\documentclass[a4paper,fleqn,usenatbib]{mnras}

\usepackage{newtxtext,newtxmath}
\usepackage[T1]{fontenc}
\usepackage{ae,aecompl}

\usepackage{graphicx}	% Including figure files
\usepackage{amsmath}	% Advanced maths commands
\usepackage{siunitx}
\usepackage{fancyvrb}

\usepackage{subfig}
\usepackage{xcolor}
\newcommand{\referee}{}
\newcommand{\refereeB}{}
\title[Chemical signatures of a warped disc]{Chemical signatures of a warped protoplanetary disc}

\author[A.~K.~Young et al.]{
Alison K. Young$^{1}$\thanks{E-mail: alison.young@leicester.ac.uk (AKY)},
Richard Alexander$^{1}$,
Catherine Walsh$^{2}$,
\newauthor
Rebecca Nealon$^{1,3,4}$,
Alice Booth$^{2,5}$
and Christophe Pinte$^{6}$
\\
$^{1}$School of Physics and Astronomy, University of Leicester, University Road, Leicester, LE1 7RH, UK\\
$^{2}$School of Physics and Astronomy, University of Leeds, Leeds, LS2 9JT, UK\\
$^{3}$Centre for Exoplanets and Habitability, University of Warwick, Coventry CV4 7AL, UK\\
$^{4}$Department of Physics, University of Warwick, Coventry CV4 7AL, UK\\
$^{5}$Leiden Observatory, Leiden University, PO Box 9513, 2300 RA Leiden, The Netherlands\\
$^{6}$Monash Centre for Astrophysics (MoCA) and School of Physics and Astronomy, Monash University, Clayton, Vic 3800, Australia
}

\date{Accepted XXX. Received YYY; in original form ZZZ}

\pubyear{2021}

\begin{document}
\label{firstpage}
\pagerange{\pageref{firstpage}--\pageref{lastpage}}
\maketitle

\begin{abstract}
Circumstellar discs may become warped or broken into distinct planes if there is a stellar or planetary companion with an orbit that is misaligned with respect to the disc. There is mounting observational evidence for protoplanetary discs with misaligned inner discs and warps that may be caused by such interactions with a previously undetected companion, giving us a tantalising indication of possible planets forming there. Hydrodynamical and radiative transfer models indicate that the temperature varies azimuthally in warped discs due to the variable angle at which the disc surface faces the star and this impacts the disc chemistry. We perform chemical modelling based on a hydrodynamical model of a protoplanetary disc with an embedded planet orbiting at a 12$^{\circ}$ inclination to the disc. {\referee Even for this small misalignment, abundances of species including CO and HCO$^+$ vary azimuthally and this results in detectable azimuthal variations in submillimetre line emission. }Azimuthal variations in line emission may therefore indicate the presence of an unseen embedded companion. Nonaxisymmetric chemical abundances should be considered when interpreting molecular line maps of warped or shadowed protoplanetary discs.
\end{abstract}


%%%%%%%%%%%%%%%%%%%%%%%%%%%%%%%%%%%%%%%%%%%%%%%%%%

%%%%%%%%%%%%%%%%% BODY OF PAPER %%%%%%%%%%%%%%%%%%

\section{Introduction}

High--resolution imaging of protoplanetary discs has opened up myriad possibilities in recent years to study the processes involved with planet formation. There are now many examples of protoplanetary discs with complex structures. In submillimetre observations of thermal emission we see features such as bright arcs, rings, gaps and spirals (e.g. \citealt{guzman2018,huang2018}), while in scattered light in addition to asymmetric features some discs have shadows (e.g. \citealt{stolker2016aa,benisty2017}). In molecular emission, observations with high spatial resolution reveal rings, `hot spots' and spiral arms (e.g. \citealt{kurtovic2018}), as well as axisymmetric extended emission.  Some of these disc structures are explained, for example the features seen in HD~142527 are due to binary--disc interaction \citep{price2018ab}, and others not. However, not yet explored is the impact of inner disc misalignment on the disc chemistry, despite inner disc misalignments being common in protoplanetary discs \citep{ansdell2020}. In this paper we present the results of modelling that predicts azimuthal variations in chemical abundances in a warped disc. This provides a new method to probe disc structure.

% Misaligned planets & warps theory
Theoretical simulations have shown that protoplanetary discs can develop warps and misaligned inner discs due to interactions with a companion \citep{papaloizou1995,facchini2013,xiang-gruess2013,nealon2018,zhu2019}, interactions with the stellar magnetic field \citep{bouvier1999} or following a flyby \citep{Xiang-Gruess2016,cuello2019,nealon2020}. Intriguingly, a warp can be induced by a planet embedded in the disc and the detection of such structures may indicate the presence of a young planet.

% Observational signatures of warps
Naturally, once it was discovered that protoplanetary discs develop warps, the question of how to detect such a structure began to be addressed. The warped structure results in azimuthal variations in the illumination of the disc by the central star, which in turn affects the disc's emission. The spectral energy distribution (SED) of a strongly warped disc has been shown to differ from that of a flat disc \citep{terquem1996,nixon2010}. However, these differences in the SED are similar to those caused by gaps or cavities in the disc. The temperature variations are predicted to produce azimuthal variations in the thermal emission, which can cause the disc to appear asymmetric around the central star \citep{nixon2010,ruge2015}.

A disc warped by the interaction with an embedded companion is likely to be accompanied by a misaligned inner disc. With a small misalignment, the inner disc may cast broad shadows on the outer disc and a highly inclined inner disc may cast narrow, radial shadows in scattered light \citep{facchini2018,nealon2019aa,zhu2019}. The warp itself may give rise to differences in the radial surface brightness profile compared to an unperturbed disc and the bright/dark sides will depend on the azimuthal position angle of the warp rather than on the inclination of the disc, as is the case for flat unperturbed discs \citep{juhasz2017}.

In addition to these continuum and scattered light features, molecular line observations also provide an indication of warped structure. The warp results in a velocity component out of the disc plane that can be detected in molecular line emission. This presents as a twist in the first moment map along the axis of the warp \citep{rosenfeld2012,casassus2015,facchini2018,zhu2019}. The different planes of the inner and outer discs are distinct in first moment maps, although the strength of the signatures depends on the inclination of the system and the azimuthal position of the warp \citep{facchini2018}. CO line observations are also more sensitive than the submillimetre continuum to temperature differences due to the high optical depth of the low rotational transitions, which means that the CO emission traces the molecular layer rather than the midplane where there is little azimuthal temperature variation. However, the contrast between illuminated and shadowed regions is less pronounced than in scattered light images \citep{facchini2018}.

The ability to measure the properties of a warp from observations would provide a valuable insight into the hydrodynamical properties of the disc. Such measurements can indicate the dynamical properties of the central object and perturber if the planet is massive (10-14 M$_{\mathrm J}$) and the disc is viscous ($\alpha = 0.15 - 0.25$) \citep{facchini2014}. In the case that the masses and orbits of the central star and companion are known, the amplitude of the warp places limits on the disc viscosity.

% Observations: PPDs that might have warps and detected embedded planets. 
Advances in high-resolution imaging in recent years have provided us with many examples of discs with features such as bright arcs and spirals, radial and azimuthal shadows and an asymmetric brightness distribution. There are currently several discs with features that may be explained by a warped structure. For example, the dipper star AA Tau hosts a disc with a twist in the HCO$^+$ moment 1 map within the inner dust ring that could be explained by a warp \citep{loomis2017}. {\referee Similarly, another dipper star 2MASS J16042165-2130284 also shows signs of a misaligned inner disc, including a twist in the moment 1 maps \citep{mayama2018} and scattered light shadows highly suggestive of a misaligned inner disc \citep{pinilla2018}.}%endref

 The morphology of the scattered light image of HD~143006 is characteristic of a warped inner disc \citep{benisty2018}. Deviations from kinematic velocities detected in CO emission from the TW Hydrae disc can be explained by a small warp \citep{rosenfeld2012} as can the observed scattered light shadows \citep{debes2017,nealon2019aa}. The velocity structure of HD~100546 detected in CO and SO lines also cannot be explained by a Keplerian rotation profile and one suggestion is that this is due to a planet interacting with the disc \citep{walsh2017,booth2018}. {\referee Two planets have been proposed to be responsible for disrupting the disc: a possible giant planet at $\sim53$~au and an inner planet at  $\sim12$~au \citep{quanz2013,quanz2015,sissa2018,brittain2019,pineda2019}. High resolution 1.3~mm and $^{12}$CO observations revealed ``wiggles'' or ``kinks'' and the Doppler flip associated with planet--disc interactions which have been attributed to the inner planet \citep{casassus2019a,PerezS2020ApJL}.}%end ref

Currently the best characterised warped protoplanetary disc system is GW Orionis. The clear scattered light structures, `twist' signature in the CO first moment map along with the tightly constrained orbital parameters \citep{kraus2020, bi2020} make this a useful reference case for observational characteristics of warps, albeit for a warm disc irradiated by three stars.
In many cases, the cause of the misalignments has not yet been identified and searches are underway for embedded planets or low mass stellar companions.

Theoretical work predicts that there is a significant azimuthal variation in temperature in a warped protoplanetary disc \citep{terquem1993,nixon2010,ruge2015}, even for small misalignments of an embedded planet {\referee \citep{juhasz2017,nealon2019aa}}, which effectively leads to a warmer and a cooler side of the disc when viewed face-on. A subsequent question is then whether or not this leads to chemical differences in such discs. If there are detectable azimuthal variations in chemical abundances this would provide another independent signature of a warp, perhaps indicative of an unseen planet, and add another dimension to line observations, reducing the ambiguity where the observations may have an alternative explanation, as appears to be the case for HD~142527 \citep{price2018ab}. Observations of molecular line emission from protoplanetary discs have not yet been exploited to the same extent as continuum observations so there is likely much still to be discovered. TW Hya has been the focus of much study and there are consequently observations of several molecular species, and these will be discussed later.

Much of the previously modelled observations of protoplanetary discs warped by an embedded {\referee planet} have exclusively considered large misalignments \citep{facchini2018,zhu2019}. However, an embedded planet has been shown to cause the disc to break even with a small ($\sim10^{\circ}$) misalignment (e.g.~\citealt{nealon2018}). The resulting misaligned inner disc casts a wide shadow across the outer disc so there is a large region of the disc with a lower temperature. The disc material therefore spends a significant period of time in the region of reduced temperature and reduced irradiation from the central star, which could impact the chemical composition and molecular line emission. Previous forays into modelling the CO emission from a warped protoplanetary disc have assumed a constant abundance or a simple `drop' model in which a depleted abundance is implemented for regions below a specified temperature (e.g. \citealt{facchini2018}). This assumption is valid for the case of a circumbinary disc where the disc temperature high enough that there is no significant freeze out of CO and little azimuthal variation in temperature. This is not true for a T Tauri star with a planetary companion that is driving the misalignment, therefore the chemical structure of the such a disc is more complex and we should consider modelling the chemical evolution.

A non--uniform chemical abundance could lead to a characteristic morphology in molecular line observations that could provide a new signature of a warped disc and a probe for embedded planets or unseen companions. Molecular species other than CO are better tracers of variations in temperature, ionization and FUV flux. Chemical modelling is essential for predicting the distributions of those molecules whose abundances are more sensitive to the physical conditions in the disc. Accurate modelling of HCN and HCO$^+$ line emission, for example, requires a large chemical network \citep{kamp2017}. In this study, we model the chemistry in a disc warped by an embedded misaligned planet to explore how the perturbed structure affects the molecular abundances in the disc. This data is then used to model the emission from selected species to determine whether this may be a probe of a warped structure and an unseen companion.

\section{Method}

{\referee
To assess the effects of non--axisymmetric physical conditions on the disc chemistry and molecular line emission we assume a simple model. }
We first run a smoothed particle hydrodynamical (SPH) model of a disc with an embedded planet on an inclined orbit to obtain the physical structure. A snapshot is selected from this model and post--processed with a radiative transfer model to calculate the temperatures. 
{\referee The chemistry is then evolved in time at each grid point under these static physical conditions.} Finally, the chemical abundances are used to compute line emission of selected molecular species. These steps are described in more detail below.
\begin{figure}
\centering
\includegraphics[width=0.95\columnwidth]{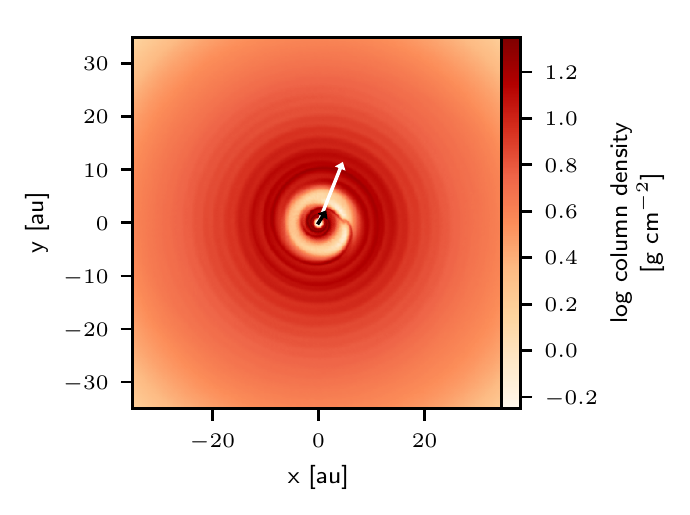}
% created with pixmapXY.py
\caption{The vertical column density of the snapshot from the hydrodynamical model, taken after 100 orbits. The planet is located at 5~au and has carved a gap, separating the inner and outer discs. The white and black arrows show the x-y projection of the angular momentum unit vectors $\hat{L}$ of the inner and outer discs respectively, scaled for readability.}
\label{fig:coldensXY}
\end{figure}

\begin{figure*}
\centering
\includegraphics[trim=0cm 0cm 0cm 0cm,clip,width=15cm]{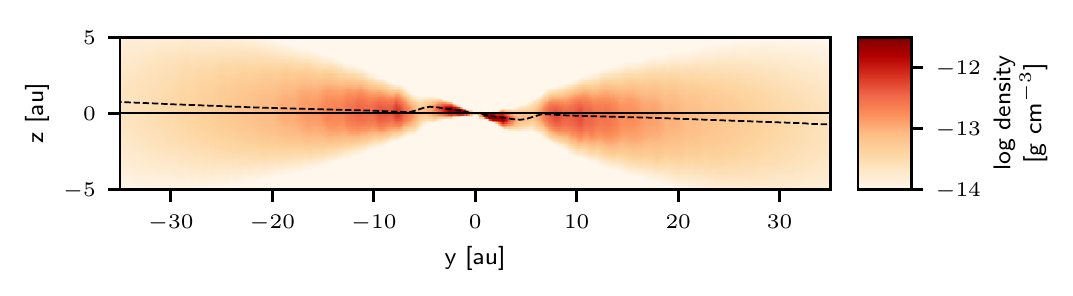}
\caption{A density cross-section of the snapshot from the hydrodynamical model. The dashed line marks the midplane of the disc in the $x=0$ plane.}
\label{fig:disc1000denssnapshot}
\end{figure*} 

\begin{figure}
\centering
\includegraphics[width=\columnwidth]{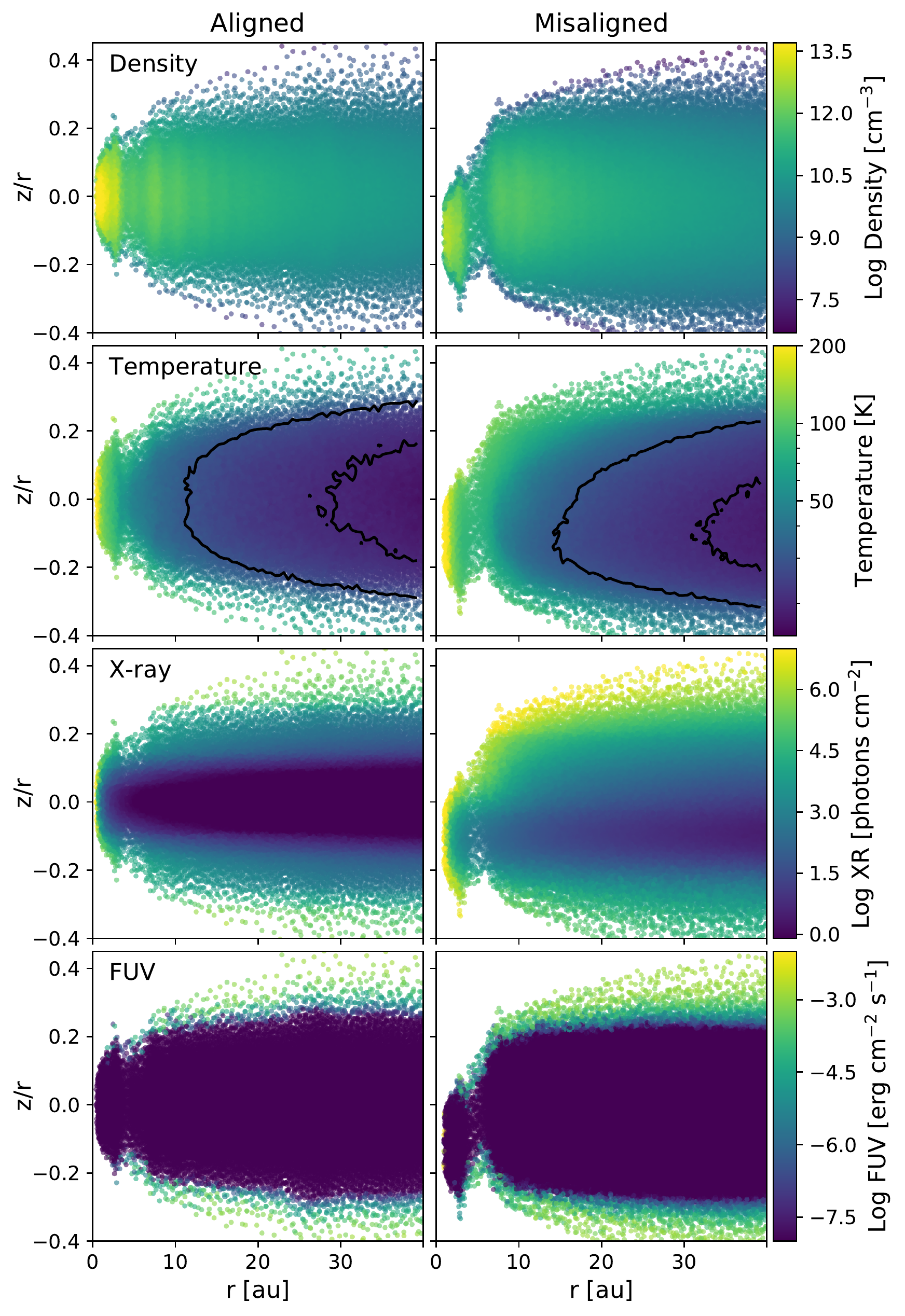}
% created with plotslicesm.py :)
\caption{Number density, temperature, X-ray field and FUV field for a cross--section of the disc. Left columns: aligned planet. Right columns: misaligned planet, slice taken along the y-axis (see Fig.~\ref{fig:disc1000denssnapshot}). Contours are drawn on the temperature panels at 20~K and 30~K to indicate the position of the CO snow surface.}
\label{fig:physicalslices}
\end{figure} 

\subsection{Hydrodynamical model}
The physical model is taken from a hydrodynamical simulation of a 6.5~M$_\mathrm{J}$ planet embedded in a disc at 5~au around a 1~M$_{\odot}$ star, which was presented in \citet{nealon2019aa}. Those calculations were conducted with {\sc phantom} \citep{price2018aa}, an SPH code that has been used extensively for simulations of warped discs.% e.g. citet{lodato2010,nixon2012...}. 
 The orbit of the planet is misaligned by 12.89$^{\circ}$ from the initial disc plane\footnote{The orbital inclination of 12.89$^{\circ}$ corresponds to an inclination of $3H/R$ at the planet's position in this disc.}. Initially, the disc extends from  0.1 to 50~au, has a mass of \num{5e-3}~M$_{\odot}$ and is modelled with \num{3.4e6} particles. The surface density profile is a power law $\Sigma(r) \propto r^{-1}$ and the viscosity is parameterised following  \citet{shakura1973}  with $\alpha_{\mathrm{ss}}=$~\num{e-3}. The disc aspect ratio is $h/r = 0.05(r/r_0)^{1/4}$, where $r$ is the cylindrical radius and $r_0 = 1$~au. The disc is assumed to be vertically isothermal with sound speed $c_{\mathrm s} = c_{\mathrm s,0}(r/r_0)^{-1/4}$. {\referee The star and planet are each represented by a sink particle of radius 0.1~au.}

A snapshot is taken after 100 orbits when the planet has carved a gap, the inner disc has separated from the outer disc and a warp has developed exterior to the orbit of the planet. There are accretion streams across the gap onto the planet and a spiral wake has developed in the outer disc. A column density plot of this snapshot is presented in Fig.~\ref{fig:coldensXY} along with the (scaled) $x-y$ projection of the average angular momentum unit vectors of the inner and outer discs, which indicate the orientation of the discs. A density cross--section along the y-axis is shown in Fig.~\ref{fig:disc1000denssnapshot}, which also shows the location of the disc midplane for this slice. Together, these figures indicate the orientation of the system. The inner disc is tilted slightly more than the outer disc, tipping away from the observer approximately along the positive $y$ direction. The outer disc tilts slightly away from the observer along a similar axis to the inner disc, with a slightly greater $x$-component, and is warped with the {\referee disc plane intersecting with the $xy$ plane approximately along the $x$-axis.} The variation in the height of the midplane in Fig.~\ref{fig:disc1000denssnapshot} indicates that the warp and misalignments of the inner and outer discs are both small.

For comparison, we modelled a second disc as a reference case with the same parameters except that the orbit of the planet is aligned with the disc and we will refer to this model as the `aligned disc'. The planet carves a gap in the disc giving rise to similar structures as for the misaligned disc except that the inner disc, outer disc and planet are co-planar. %These figures are in the Appendix.

\subsection{Radiative transfer: temperature calculation}
\label{sec:RT}
Dust temperatures were calculated with {\sc mcfost} \citep{pinte2006,pinte2009}. The stellar luminosity\footnote{We note that this is written incorrectly in \citet{nealon2019aa}. The value quoted here is what was used for the calculations.} is 2.1~L$_{\odot}$, taken from the 1Myr Siess isochrone \citep{siess2000},  and the effective temperature is 4278~K.  {\sc mcfost} maps the particles from the hydrodynamical model snapshot onto a 3-D Voronoi mesh to perform the radiative transfer calculations. Before creating the mesh, particles with $r<0.9$~au are removed to prevent unphysical shadowing due to the low particle numbers in the inner cavity. To prevent similar effects from extremely irregularly--shaped cells on the edge of the disc, the density is reduced such that they are optically thin.

It is the Voronoi cells, rather than SPH particles, that we consider for calculating the X-ray and far ultraviolet (FUV) fields and for the chemical calculations. 
The size distribution follows the model of \citet{1977mrn}, with grain sizes 0.03~$\mu$m to 1~mm and power law exponent of 3.5. The dust grain model is the `smoothed astronomical silicate' \citep{drainelee1984,laor1993,weingartner2001}. We assume that dust and gas temperatures are equal and discuss this assumption further in section~\ref{sec:timescales}.

 \subsection{X-ray and FUV fields}
 The X-ray field was calculated following the formulation of \citet{krolik1983}, assuming the star to be a classical T Tauri star. We used the binned X-ray spectrum of \citet{nomura2007} which was created by fitting the observed spectrum of TW Hydrae with a two temperature thermal thin plasma model. The X-ray luminosity is L$_\mathrm{X} = $~\num{1.6e30}~erg~s$^{-1}$. The energy--dependent optical depth is
\begin{equation}
  \tau (\epsilon) = \frac{\mathrm{N}_\mathrm{H}}{ 4.4\times 10^{21} \mathrm{cm}^{-2}  } \left(\epsilon/ 1 \mathrm{keV}\right)^{-\alpha}.
 \label{eq:xraytau}
\end{equation}
  The constant $\alpha=2.485$ \citep{glassgold1997,glassgold2000} and $\epsilon$ is the photon energy. The column number density of hydrogen nuclei, N$_\mathrm{H}$, is calculated from the line-of-sight mass column density from the cell to the star, given by {\sc mcfost}. The frequency--dependent attenuating factor is 
  
 \begin{equation}
 J(x,\tau) = x^{-\alpha} e^{-x - \tau x^{-\alpha}},
 \end{equation}

with $x = \epsilon/1\mathrm{keV}$. The attenuated spectrum in the disc is then
\begin{equation}
I(\epsilon,x,\tau)_{\mathrm{attenuated}}= I(\epsilon)J(x,\tau) .
\end{equation}
The X-ray field for a slice through the disc is shown in Fig.~\ref{fig:physicalslices}.

The continuum FUV field in the disc is calculated considering the contribution of both the star and the external radiation field. The stellar FUV luminosity is assumed to be $L_{\mathrm{fuv} }=10^{-3} L_{\mathrm{star}}$ and the external radiation field is the Habing field $G_0$. The optical depth is $\tau_{\mathrm{fuv}} = 8\times 10^{-22} \times n_{\mathrm H}$ \citep{adams2004}, with the column density, $n_{\mathrm H}$, calculated line-of-sight to the star and vertically through the disc respectively for the stellar and external contributions. {\referee We assume that the dust is well mixed. Since the UV absorption opacity is dominated by small dust grains, this assumption holds for most of the lifetime of the disc. }

\subsection{Chemical model}
Chemical abundances are calculated for the cells in the Voronoi grid created by {\sc mcfost}. The chemical model comprises a gas-phase network and gas-grain reactions and is the same as that of \citet{walsh2015}. The gas phase network is based on the `RATE12' network from the UMIST database for astrochemistry \citep{mcelroy2013}. This network includes gas--phase two--body reactions, direct cosmic ray ionization, cosmic ray induced photodissociation and ionization, FUV driven photodissociation and photoionisation and cation--grain recombination. Direct X-ray ionization and ionization--induced X-ray reactions are also included. The X-ray ionization rate at each point in the grid is calculated using the X-ray spectrum at that point and the elemental composition (see \citealt{walsh2012}). {\referee Photodissociation rates are calculated from the wavelength--integrated UV flux at each grid cell. This includes internally--generated FUV photons from cosmic ray photon interactions. The photodissociation and photoionisation rates are calculated for the \citet{draine1978aa} field extended into the near-UV \citep{vandishoeck1982}. Self-shielding is taken into account for H$_2$, CO and N$_2$ using tabulated shielding functions\footnote{These tabulated values are available at \url{https://home.strw.leidenuniv.nl/~ewine/photo/shielding_functions.html}.} and the estimated column density of these species to the star, as described in detail in \citet{walsh2015}. We note that Lyman~$\alpha$ is not included in the models presented here.}%end ref

We consider adsorption of gas--phase species onto dust grains, thermal desorption, cosmic--ray induced desorption and photodesorption. The molecular binding energies are taken from the `RATE12' network \citep{mcelroy2013} with the updated values for HCN \citep{noble2013}. Dust grains are assumed to be spherical with a radius of 0.1~$\mathrm{\mu}$m, \num{e6} surface binding sites and a number density of \num{e-12} relative to the gas number density. H$_2$ formation is included by assuming the rate of production is half the rate of the adsorption of H onto grains.

Initial abundances were calculated for a molecular cloud environment with $T_{\mathrm{ gas}}=T_{\mathrm {dust}}=10$~K, $n=$~\num{e6}~cm$^{-3}$ and A$_{\mathrm{v}}= 10$~mag. The abundances were evolved for \num{e5} years to simulate the formation of molecules, such as water ice, that occurs prior to disc formation. The elemental abundances can be found in Table~\ref{tab:abunds} and selected molecular abundances calculated from the elemental abundances and used to initialise the disc chemistry calculation can be found in Table~\ref{tab:molabunds}.

The chemistry calculation was performed for each Voronoi cell, using the temperatures calculated by {\sc mcfost} and the X-ray and FUV fields estimated as described above. The chemical abundances were extracted after evolving under the fixed physical conditions for 1~Myr. We choose this timescale because here we are evolving the chemistry from molecular cloud abundances to the age of the disc, assumed to be 1~Myr. {\referee We discuss the chemical timescales further in \ref{sec:mainchemtimes}}. 

\begin{table}
\centering
\caption{Elemental abundances for the chemical model. These are the low--metal abundances of \citet{graedel1982aa} except where otherwise indicated. $a(b)$ means $a\times 10^b$.}
\label{tab:abunds}
\begin{tabular}{ll}
Element & Abundance (n$_{\mathrm j}/n_{\mathrm{H}}$) \\
\hline
H$_2$  & $0.5$ \\
He   & $9.75(-2)$ \\
N    & $7.5(-5)$ $^\mathrm{a}$ \\ %cardelli91
O    & $3.2(-4)$ $^\mathrm{b}$ \\ % meyer1998
C$^+$   & $1.4(-4)$ $^\mathrm{c}$ \\ %cardelli1996
S$^+$   & $8(-8)$   \\
Si$^+$  & $8(-9)$   \\
Fe$^+$  & $3(-9)$   \\
Na$^+$  & $2(-9)$   \\
Mg$^+$  & $7(-9)$   \\
P$^+$   & $3(-9)$  \\
Cl$^+$  & $4(-9)$ \\
\end{tabular}

\raggedright
$^\mathrm{a}$ \citet{cardelli1991}\\
$^\mathrm{b}$ \citet{meyer1998}\\
$^\mathrm{c}$ \citet{cardelli1996}
\end{table}

\begin{table}
\centering
\caption{Selected molecular abundances used to initialise the disc chemistry model. These were obtained after evolving the elemental abundances under the physical conditions of a molecular cloud  ($T_{\mathrm{ gas}}=T_{\mathrm {dust}}=10$~K, $n=$~\num{e6}~cm$^{-3}$ and A$_{\mathrm{v}}= 10$~mag) for \num{e5} years.  $a(b)$ means $a\times 10^b$.}
\label{tab:molabunds}
\begin{tabular}{ll}
Species & Abundance ($n_{\mathrm j}/n_{\mathrm{H}}$) \\
\hline
CN & $4.8(-11)$ \\
CO  & $9.5(-9)$ \\
CO ice  & $7.1(-6)$ \\
CS & $6.0(-9)$ \\
HCO$^+$  & $3.0(-12)$ \\
HCN & $1.7(-11)$\\
N$_2$H$^+$ & $1.7(-11)$ \\
SO & $6.0(-11)$ \\
H$_2$O & $1.5(-6)$ \\
H$_2$O ice & $4.8(-5)$ \\
\end{tabular}
\end{table}

\subsection{Radiative transfer: synthetic line maps}
The calculated chemical abundances were used to model synthetic line maps with {\sc mcfost} using the same setup as for calculating the temperatures outlined in section~\ref{sec:RT} . Additionally, the velocity data from the hydrodynamical model is passed to {\sc mcfost}. Abundances are read in for each Voronoi cell from the chemistry calculation. For the $^{13}$CO abundance we assume a $^{12}\mathrm{CO}/^{13}\mathrm{CO}$ ratio of 77 after \citet{woods2009}, though their work and that of \citet{miotello2014} demonstrates that some fractionation of CO isotopologues is to be expected. We assume a  $^{12}\mathrm{CO}/\mathrm{C}^{18}\mathrm{O}$ ratio of 550 \citep{wilson1994aa} but this may overestimate the line intensity of C$^{18}$O \citep{williams2014}. {\referee We also model the line emission assuming constant molecular abundances to compare with the results from the chemical modelling. For those constant abundance models: $n_{\mathrm{CO}}/n_{\mathrm H} = 10^{-4}$, $n_{\mathrm{HCN}}/n_{\mathrm H} = 10^{-9}$ and $n_{\mathrm{SO}}/n_{\mathrm H} = 5\times10^{-9}$.} Molecular line transition data was taken from the LAMDA database \citep{schoier2005aa}. Level populations are calculated assuming local thermodynamic equilibrium and ray--tracing radiative transfer is implemented. The local line profile $\phi$ as a function of velocity $v$ is
\begin{equation}
\phi(v) = \frac{c}{\sqrt{\pi \sigma^2}} \exp{\left(-v^2/\sigma^2\right)},
\end{equation}
where
\begin{equation}
\sigma^2 = \frac{2k_B T}{\mu m_{\mathrm p}} + v_{\mathrm {turb}}^2.
\end{equation}
Here $c$ is the speed of light, $k_B$ is the Boltzmann constant, $\mu m_{\mathrm p}$ is the molecule mass and the turbulent velocity $v_{\mathrm {turb}}= 0.05$~km~s$^{-1}$.

For the face--on line maps we model emission between $\pm 10$~km~s$^{-1}$ with channels at intervals of 0.5~km~s$^{-1}$ and for inclined line maps we extend the velocity range to $\pm 30$~km~s$^{-1}$ with  0.25~km~s$^{-1}$ channel intervals. This ensured there were several line--free channels for performing the continuum subtraction. The line maps from {\sc mcfost} were smoothed spatially with a Gaussian beam, Hanning smoothing was applied, and the continuum was subtracted in the image domain using the {\sc casa}\VerbatimFootnotes
\footnote{{\referee The tools used were \Verb"imsmooth", \Verb"imcontsub" and \Verb"specsmooth".}}
 software package \citep{mcmullin2007aa}.
Moment 0 (integrated intensity) maps were created by integrating over the full line width to where the line intensity fell to zero.

\section{Results}

{\referee We first compare the calculated chemical abundances with the results of prior chemical models of T Tauri discs. We then discuss the differences in chemical abundances found for the warped disc and present synthetic moment 0 maps of the simulated line emission. Finally, we consider the validity of our simplified model and comment on the best tracers of warped disc structures. }

\subsection{Model validation}

{\referee
\subsubsection{Chemical abundances}
We have verified that the chemical abundances calculated here are in line with previous models of T Tauri discs. Abundances of CO and HCN from the aligned model (Fig~\ref{fig:chemslices}) match well with the base chemical model of \citet{kamp2017}. The CO column density is in excellent agreement with the models presented by \citet{wakelam2019}. The column density of SO at 10~au on the illuminated side of the disc (Fig.~\ref{fig:coldensslices}, 90$^{\circ}$) is also in agreement with \citet{wakelam2019} but decreases less steeply with radius. The column density of HCN in the models of \citet{wakelam2019} is slightly lower. The other consideration here is the height above the midplane to which the model extends. In their model of a T Tauri disc, \citet{walsh2015} find an HCN abundance $\gtrsim$\num{e-9} at $z/r>0.25$ at 10~au, although \citet{kamp2017} found the HCN abundance to be negligible for $z/r>0.3$ at the same radius. Due to the limitations of SPH resolution we are only able to model the disc for $z/r<0.3$ at 10~au. It is therefore possible that we miss the contribution of HCN above the main molecular layer and that the column density of HCN should be higher than we obtain here. C$_2$H abundances for $|z/r| < 0.3$ are similar to those from the model of \citet{cleeves2018} but similar to HCN, the C$_2$H layer higher in the disc may not be fully resolved.  
}

\subsection{Chemical differences between aligned and misaligned discs}
\begin{figure}
\centering
\includegraphics[width=\columnwidth]{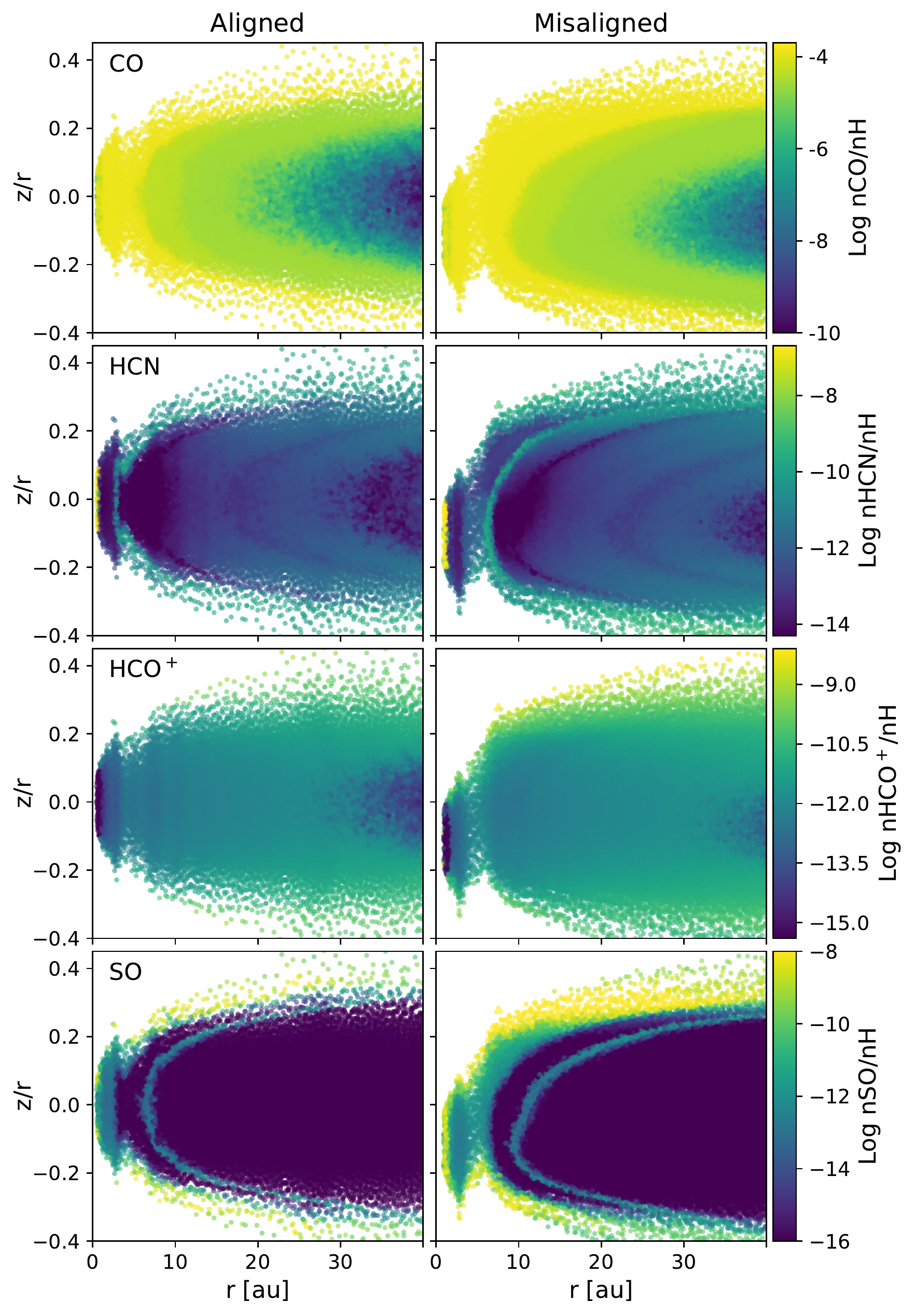}
\caption{Selected molecular abundances for the same disc cross--sections as in Fig.~\ref{fig:physicalslices}. Left columns: aligned planet located at 5~au. Right columns: misaligned planet.}
\label{fig:chemslices}
\end{figure}

Cross--sections of the discs with an aligned and misaligned planet showing the abundances of four species after $10^6$~years are presented in Fig.~\ref{fig:chemslices}. For CO and HCN, the abundance near the midplane decreases more slowly as a function of radius for the warped disc. The abundance of CO remains $>10^{-5}$ to a greater depth for the misaligned case, which we attribute to the higher temperatures (see Fig.~\ref{fig:physicalslices}).
All four species show asymmetrical abundances on upper and lower sides of the disc. The abundances of SO and HCO$^+$ are around an order of magnitude higher in the upper surface layer of the disc.

\begin{figure*}
\centering
\includegraphics[width=0.75\textwidth, trim=0.5cm 4cm 1.5cm 1cm,clip]{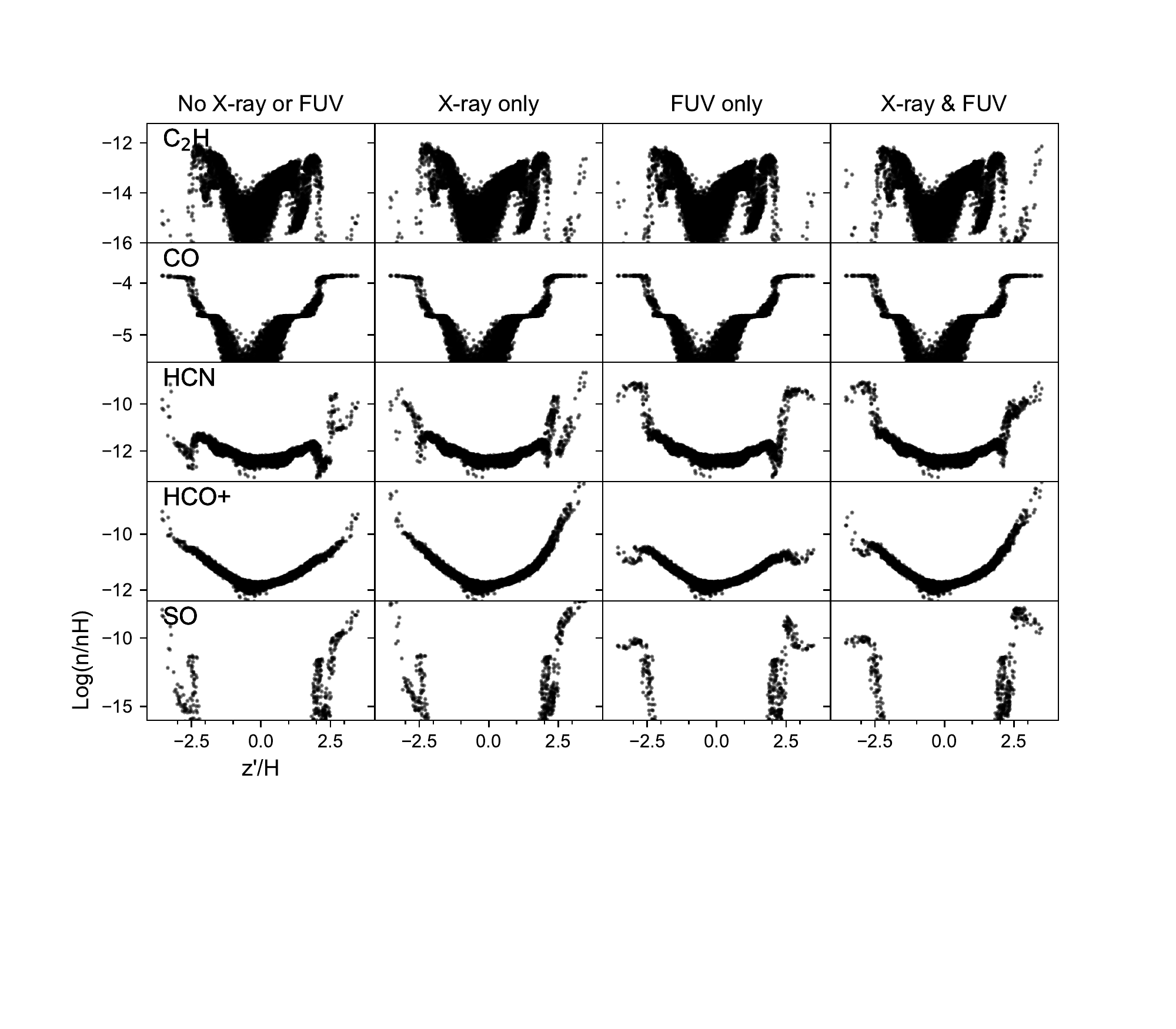}
\caption{A comparison of the vertical distribution of abundances of selected species at 30~ au for chemical models with and without X-ray-- and FUV--driven reactions. Abundances were calculated for an $r-z$ slice of the disc, taken along the $y$-axis of Fig.~\ref{fig:coldensXY}. The height from the midplane in scale heights, $z'/h$, is defined from the midplane of the warped disc. }
\label{fig:fuvxrcompare}
\end{figure*}

To clarify the origin of these asymmetries, the chemical calculations were repeated switching off FUV and X-ray driven reactions and the resulting abundances for the warped disc at a radius of 30~au are presented in Fig.~\ref{fig:fuvxrcompare}. The abundances are plotted as a function of scale height $z'/h$, where $z'$ is the distance perpendicular to the midplane (c.f. Fig.~\ref{fig:disc1000denssnapshot}).
%The height above the warped midplane, $z'$, is calculated from the particle angular momentum vectors. 
We would expect to see differences between the abundance profiles where X-ray or FUV chemistry contributes significantly to the molecular abundances. In the regions of the disc studied, the CO abundance is driven by thermal freeze out and desorption reactions but other species including HCO$^+$ are influenced by X-ray driven reactions near to the disc surface.

{\referee
The abundance of HCO$^+$ in the upper disc layers is increased by a factor of $\sim 100$ by the inclusion of X-ray driven reactions compared to only gas--phase chemical reactions. The upper side of the warped disc is more exposed to stellar X-ray photons (c.f. Fig.~\ref{fig:physicalslices}). This increases the ionisation rate of He, which increases the production of HCO$^+$ on the brighter side of the disc via the following reactions: 
 
 \begin{equation}
 \begin{array}{l}
 \mathrm{He} + \text{X-ray} \longrightarrow \mathrm{He}^+ + \mathrm{e}^-\\ 
 \mathrm{H}_2 + \mathrm{He}^+ \longrightarrow \mathrm{H}_2^+ +\mathrm{He}\\
 \mathrm{H}_2^+ + \mathrm{H}_2 \longrightarrow \mathrm{H}_3^+ +\mathrm{H}\\
 \mathrm{H}_3^+ + \mathrm{CO} \longrightarrow \mathrm{HCO}^+ + \mathrm{H}_2
 \end{array}
\end{equation}

HCN abundances are also increased by X-ray driven reactions. After $10^6$~ years, HCN has formed and adsorbed onto the dust grains at $r\gtrsim2$~au. In most of the studied regions of the disc, the main origin of gas--phase HCN is the photodesorption of HCN ice. This is driven by both secondary X-ray photons and direct FUV photons. FUV reactions contribute to the HCN abundance close to the surface of the disc but the main effect is due to X-ray photons since X-rays penetrate deeper into the disc. SO abundances are mainly temperature--driven. Close to the surface, X-ray reactions act to increase the SO abundance through secondary photons which induce the photodesorption of SO from the ice phase and photodissociation reactions reduce the abundance. }% check this section

\begin{figure*}
% current figure made with plotmolandgascd.py in above folder
% raise to 300dpi before submitting
	\centering
	\includegraphics[width=0.9\textwidth]{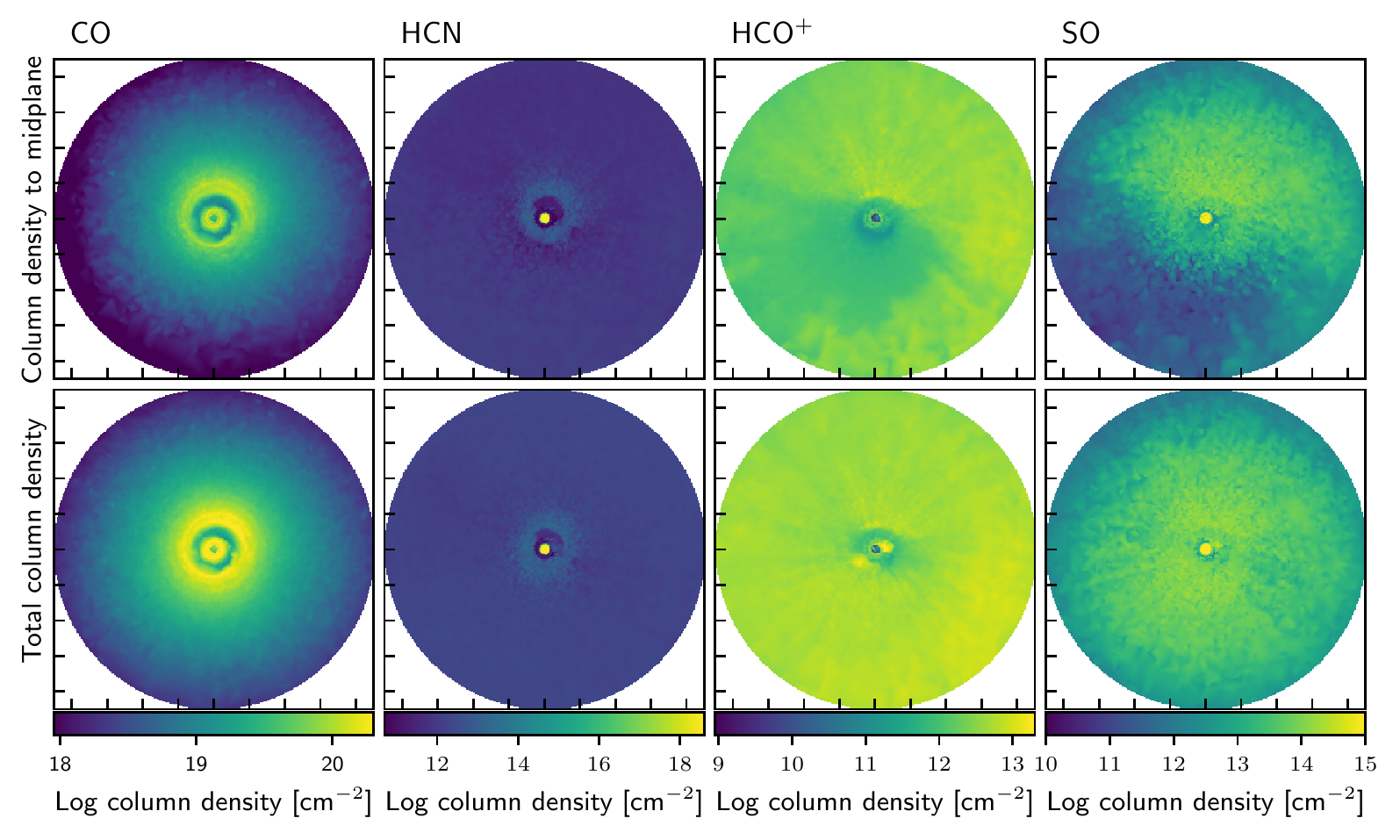}
	\caption{Column density to the warped midplane (upper panels, see Fig.~\ref{fig:disc1000denssnapshot}) and total column density (lower panels) of selected species after 1~Myr. Tick marks are at 10~au intervals.}
	\label{fig:coldensspatial}
\end{figure*}

\subsection{Chemical asymmetries}

\begin{figure*}
	\centering
	\includegraphics[width=\textwidth]{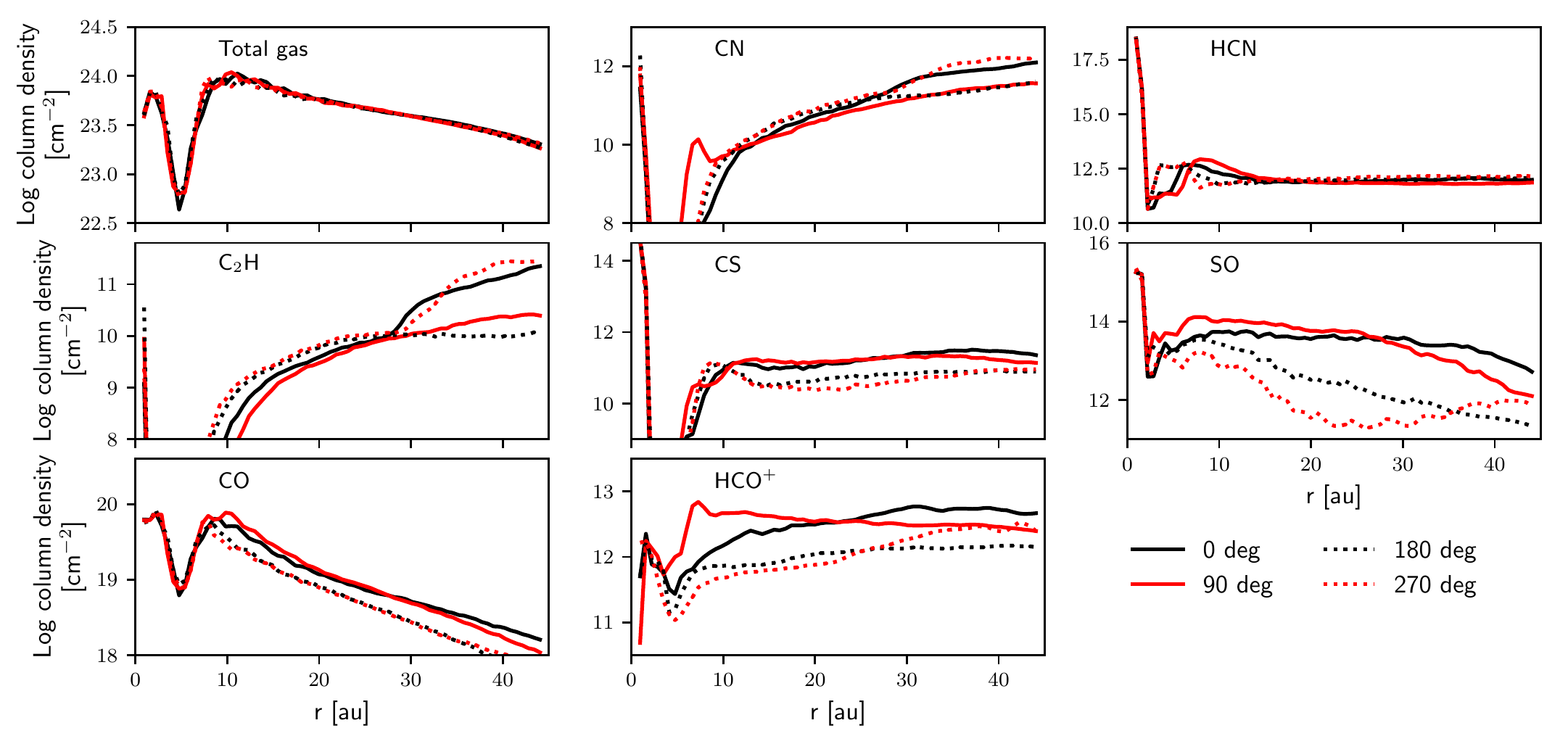}
	\caption{The column density down to the warped midplane (i.e. this is not the total column density of the disc) of selected species after 1~Myr averaged in slices 90$^{\circ}$ apart, where the positive $x$-axis is $0^{\circ}$ and progressing anti-clockwise.}
	\label{fig:coldensslices}
\end{figure*}

We calculate column densities from the top of the disc down to the midplane to simulate the column of gas that would contribute to observed emission if the midplane is optically thick. We do this first to look for asymmetry that could be detectable before later performing the radiative transfer modelling. We obtain the vertical column density from {\sc mcfost} for particles close to the (warped) midplane. The upper and lower sides of the disc are roughly symmetrical about the line of nodes ($\sim$ along the x-axis) which means the total column density conceals much of the asymmetry in the chemistry. The spatial distribution of selected species is shown in Fig.~\ref{fig:coldensspatial}. Fig.~\ref{fig:coldensslices} shows radial variation of the averaged column density for selected species at slices taken at 90$^{\circ}$ intervals anticlockwise, starting along the positive x-axis (0$^{\circ}$). The apparent differences in total gas column density within the gap between inner and outer disc are caused by the accretion stream onto the planet. Between 10 and 20~au, the oscillations in column density trace the spiral wake of the planet. The column densities of several species show substantial azimuthal variation, which we now discuss.

\begin{figure}
\centering
%made in ipynb tau=1plots
	\includegraphics[width=0.9\columnwidth]{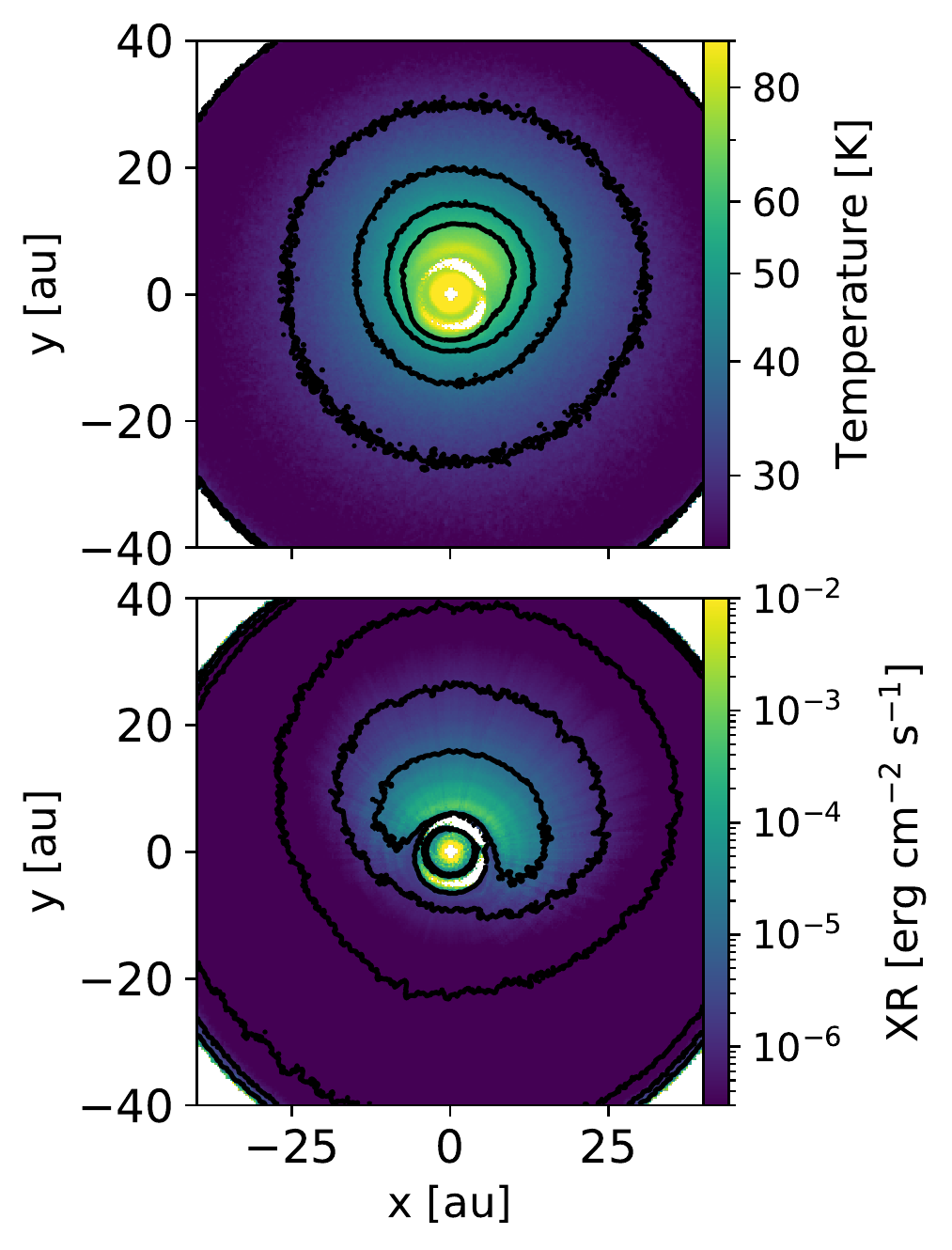}
	\caption{Top: Temperature of the warped disc at the  $\tau(70\mu\mathrm{m})=1$ surface. Contours are drawn at 30, 40, 50 and 60~K. Bottom: X-ray illumination of the disc at the $\tau(70\mu\mathrm{m})=1$ surface. Contours are drawn at \num{e-7}, \num{e-6} and \num{e-5} erg~cm$^{-2}$~s$^{-1}$.}
	\label{fig:XRtau1}
\end{figure}

{\referee
The column density of SO has the largest azimuthal variation with a difference of up to a factor of 100 for opposite sides of the disc from this model. In Fig.~\ref{fig:XRtau1}, lower panel, we show the X-ray illumination of the disc by plotting the X-ray field at the $\tau(70\mathrm{\mu m})=1$ surface (which is at $z=$~1.5-2.5 au at $r=20$~au). The X-ray field is strongly affected by shadowing due to the inner disc and therefore the X-ray illumination is similar to that seen in the synthetic scattered light image of the model that was presented in \citet{nealon2019aa}. The azimuthal temperature variation is more symmetric but still decreases faster with radius on one side of the disc. We saw in Fig.~\ref{fig:fuvxrcompare} that the HCO$^+$ abundance is strongly driven by X-rays. The spatial distribution of the HCO$^+$ abundance can thus be explained by the asymmetric X-ray irradiation.

The azimuthal difference in column density is smaller for CO, but enough to cause the position of the snow line to vary azimuthally by a few au. HCN appears to trace the inner disc and the inner rim of the outer disc, which corresponds to the arcs of increased X-ray illumination on opposite sides of the outer disc inner rim.

The abundance of CN is low in the inner part of the disc modelled here, however there is a nonaxisymmetric distribution and at larger radii CN emission may show evidence of the misaligned inner disc and warped outer disc structures. Similarly, the column density of C$_2$H diverges sharply from 30~au on opposite sides of the disc and also increases with radius. This indicates that C$_2$H may be a useful indicator of a warped disc at larger radii than considered here. 
}

\subsection{Synthetic Moment 0 maps}
%*explain choice of molecules down to which have been detected*
\begin{figure}
\centering
\includegraphics[width=0.9\columnwidth]{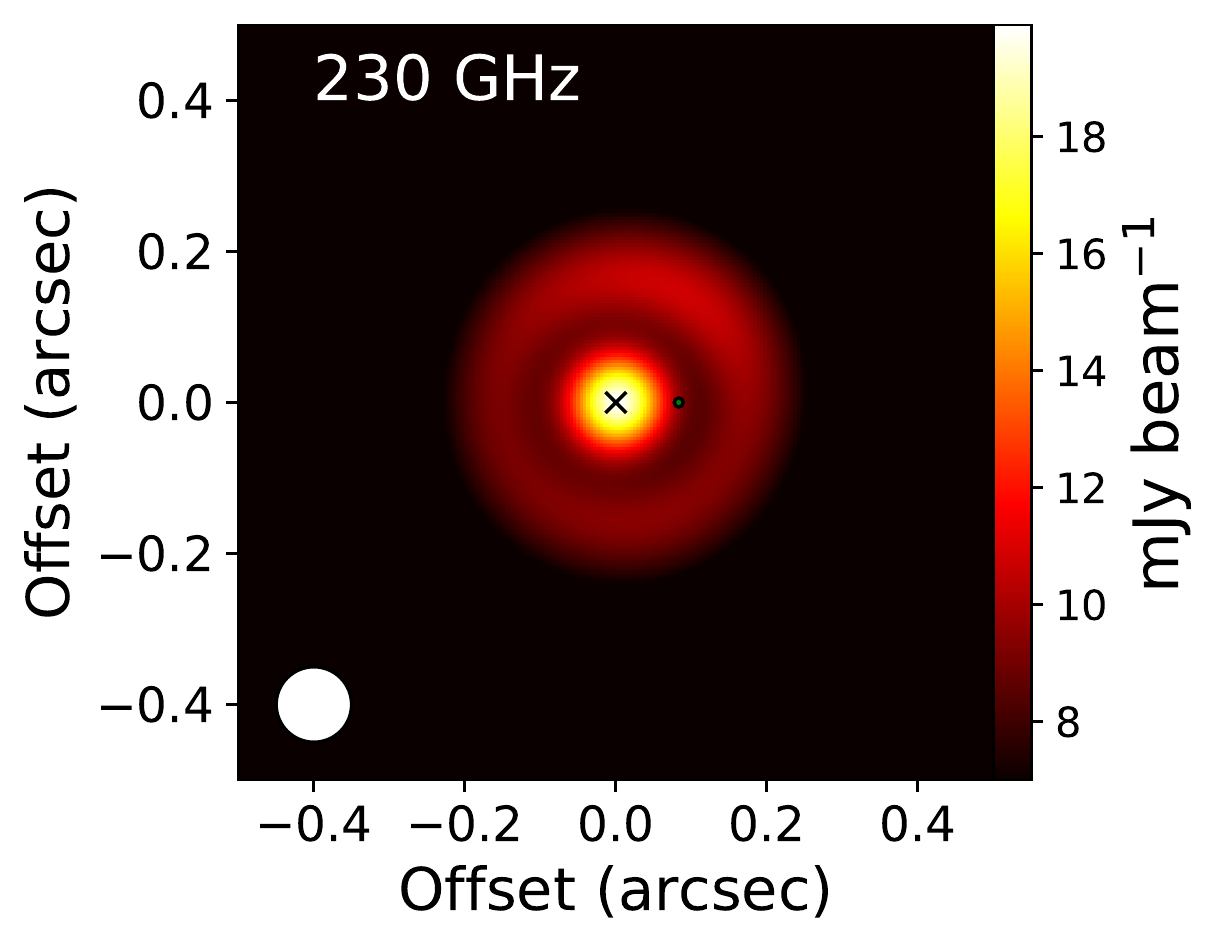}
\caption{230~GHz continuum image smoothed with a 0.1 arcsec Gaussian beam. The positions of the star and planet are marked with an `x' and a green dot respectively. With linear scaling, the inner disc is bright and the inner rim of outer disc appears as a ring. There is a brighter arc coinciding with the warp, where the temperature is higher.}
\label{fig:continuum}
\end{figure}
{\referee We have shown that the warped structure gives rise to asymmetry in the chemical abundances. Next, we determine whether this produces observable effects in the molecular line emission.} Before discussing the molecular emission, the 230~GHz continuum (the frequency of the CO~$(2-1)$ transition) is shown for comparison in Fig.~\ref{fig:continuum}. The warm inner disc is bright and there is a brighter arc on the inner edge of the outer disc at the position of the warp due to the increase in temperature there. The gap is clear but there are no further indications of the structure. As for the total column density in Fig.~\ref{fig:coldensslices}, the structure is nearly axisymmetric. Of course, there could be additional dust trapping processes not considered here that could lead to stronger asymmetries in the outer disc in the dust continuum. From the continuum image alone, it is not possible to deduce the cause of the asymmetry here.

\begin{figure*}
\includegraphics[width=\textwidth,trim=6cm 7.8cm 5.5cm 1cm, clip]{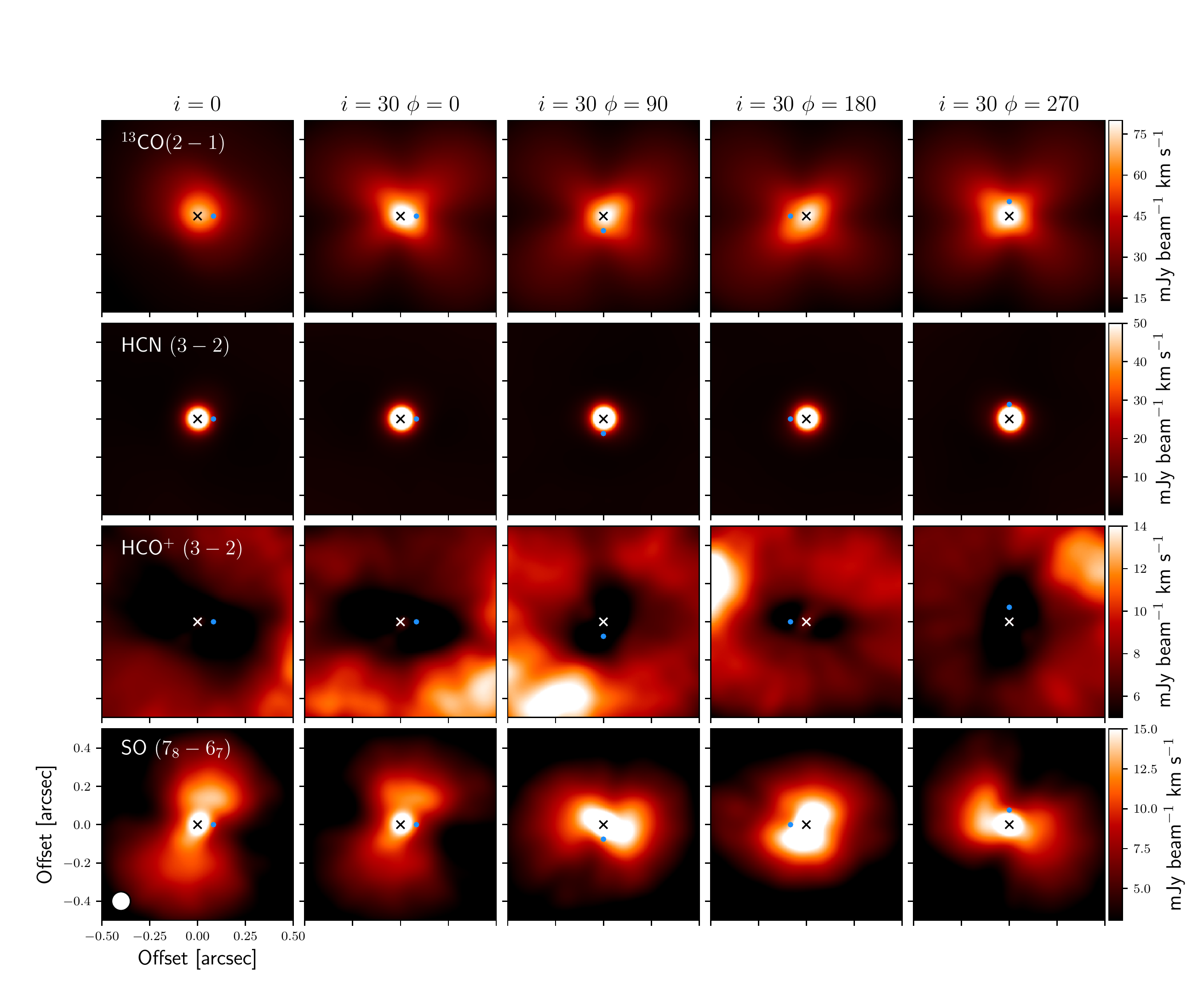}
% from /Users/akg28/Documents/WarpChem/RadTrans mom0grid.py
\caption{Moment 0 (integrated intensity maps) for selected lines observed from a distance of 60~pc and  smoothed with a 0.1 arcsec Gaussian beam. Each image spans $60 \times 60$~au. The position of the star is marked  with an `x' , the planet is marked with a blue dot and the beam size is shown by the white circle. The left hand column shows the face-on, $i=0^{\circ}$, view and the remaining columns show the lines with the disc rotated by four angles ($\phi$) and then viewed at $i=30^{\circ}$.}
\label{fig:mom0maps}
\end{figure*}

{\referee
Continuum--subtracted moment 0 (integrated intensity) maps for selected species and transitions are presented in Fig.~\ref{fig:mom0maps}. Each of these molecular lines presents different spatial asymmetry and on different scales. $^{13}$CO~$(2-1)$ emission is non--axisymmetric on the scale of tens of au. The HCN $(4-3)$ map shows a central emission peak. HCO$^+$ emission forms a crescent at $i=0$. As well as the central peak of SO emission, there is a `bow-tie' of emission aligned with the axis of the warp. The emission from the upper section is brighter, which is consistent with the higher SO abundance in that part of the disc on the side facing the observer. SO emission from protoplanetary discs is faint but, where it is detectable, this asymmetry should also be measurable. CS $(5-4)$ and CN $(3_{7/2}-2_{5/2})$ were also modelled but the emission is faint on these scales, tracing mainly the inner disc like HCN, and these are not shown here.}

We also consider the emission morphology at a viewing inclination of 30$^{\circ}$. The disc structure is nonaxisymmetric so we rotate the disc by $\phi =$ 0, 90, 180 and 270$^\circ$ in the $x-y$ plane before altering the viewing inclination to examine the difference caused by the azimuthal position of the planet and warp. These moment 0 maps are also presented in Fig.~\ref{fig:mom0maps}.

The asymmetry in the CO emission is more subtle at $i=30^\circ$ but is nonetheless present with the extent of the emission greater on one half of the disc. This asymmetry is seen more clearly in the azimuthal brightness profiles of selected $^{13}$CO~$(2-1)$ maps, presented in Fig.~\ref{fig:aza}. The drop in brightness is $\sim$~10 per cent greater for the misaligned disc and the profile shows a large drop in brightness spanning approximately half of the disc in comparison to the multiple peaks in the profile of the aligned disc.

The HCN emission does not vary noticeably at $i=30^{\circ}$, still showing a central emission peak. The morphology of the SO emission is more variable with the azimuthal position of the warp at $i=30^\circ$. At some values of $\phi$, the emission within $r \lesssim 20$~au forms an irregular elongated clump. At larger scales, the `bow-tie' shape is still present at $\phi=0$ and 270$^{\circ}$.

The HCO$^+$ moment 0 map has a distinctive bright spot or arc at a radius of $\sim 30$~au, seen even when viewed at moderate inclination (Fig.~\ref{fig:mom0maps}). This corresponds to the region where the HCO$^+$ column density is highest (outside of the central regions) and indicates that a planet at $\sim 5$~au can affect the line emission at much greater radii. This strong asymmetry should easily be apparent in observations in which HCO$^+$ is detected. The HCO$^+$ abundance is very sensitive to the stellar X-ray irradiation (see e.g. \citealt{teague2015}). Variations in the height of the disc such as from a planet--induced spiral wake may therefore also cause sufficient shadowing to affect the morphology of HCO$^+$ emission. This means that this HCO$^+$ morphology may not uniquely trace warps.

{\referee
\begin{figure*}
\subfloat[a][]{
{\includegraphics[width=0.43\textwidth]{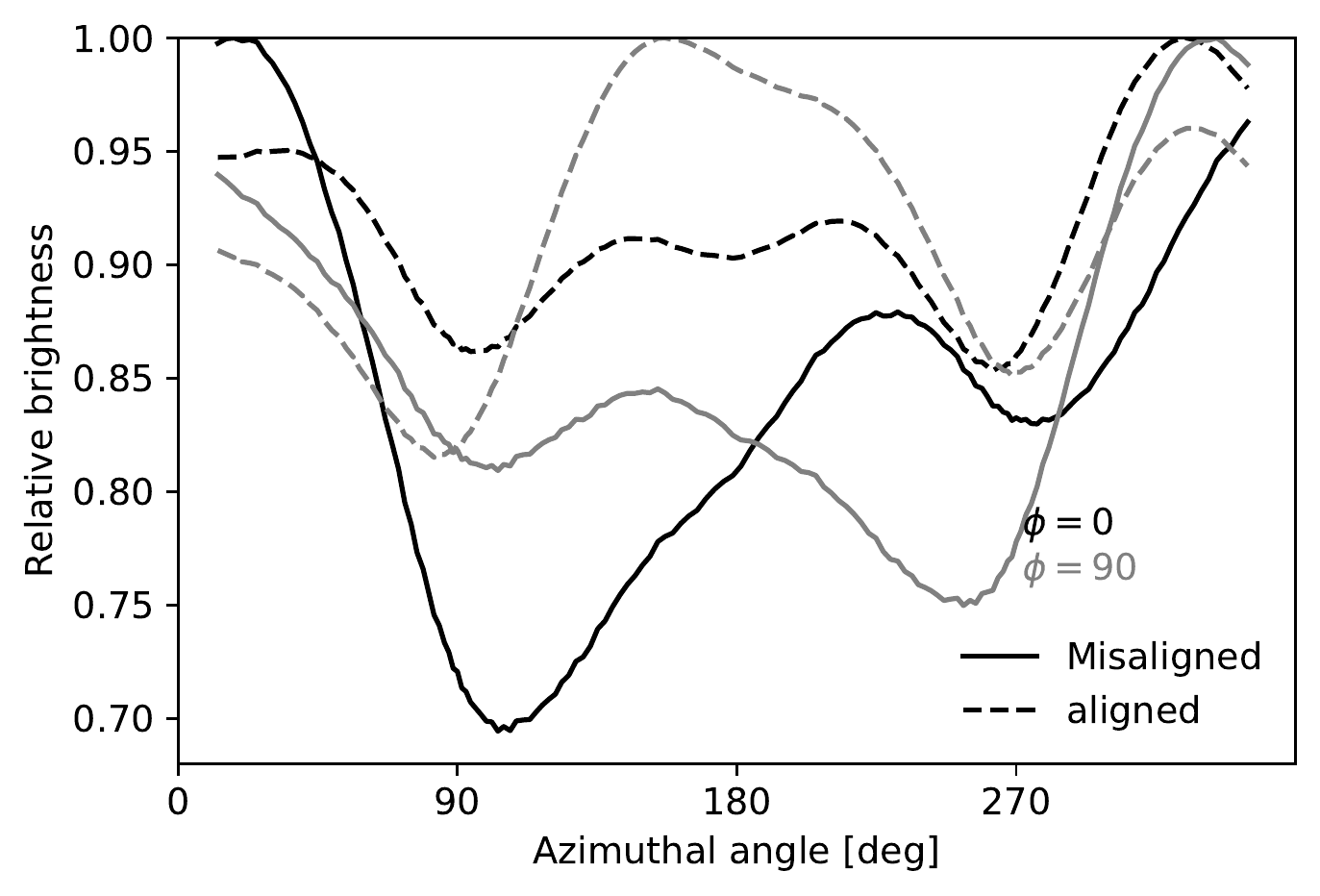}}% made with ipynb azbrightness
\label{fig:aza}
}
\qquad
\subfloat[b][]{
\includegraphics[width=0.43\textwidth]{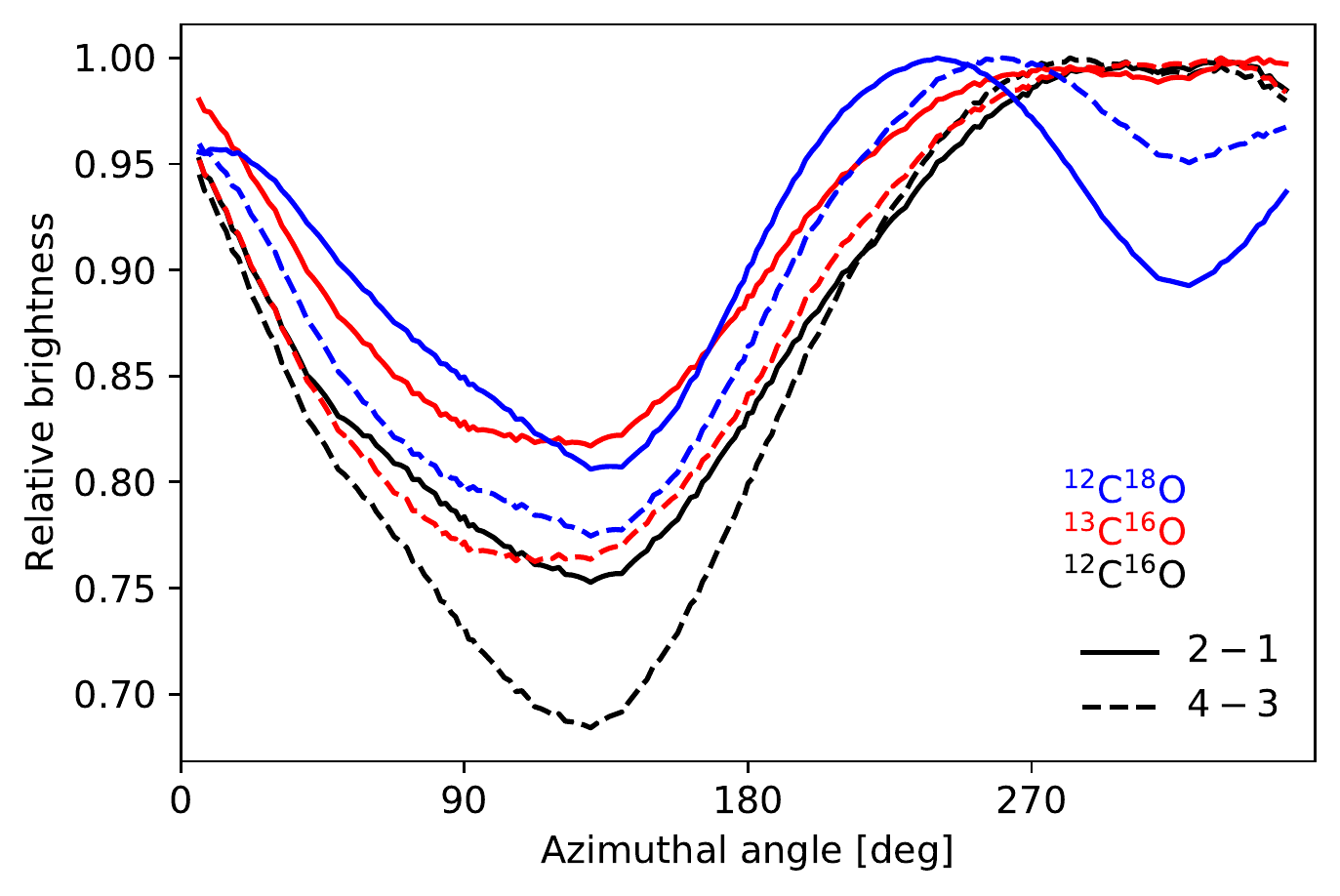} % made with ipynb azbrightness
\label{fig:azb}
}
\\
\subfloat[c][]{
\includegraphics[width=0.43\textwidth]{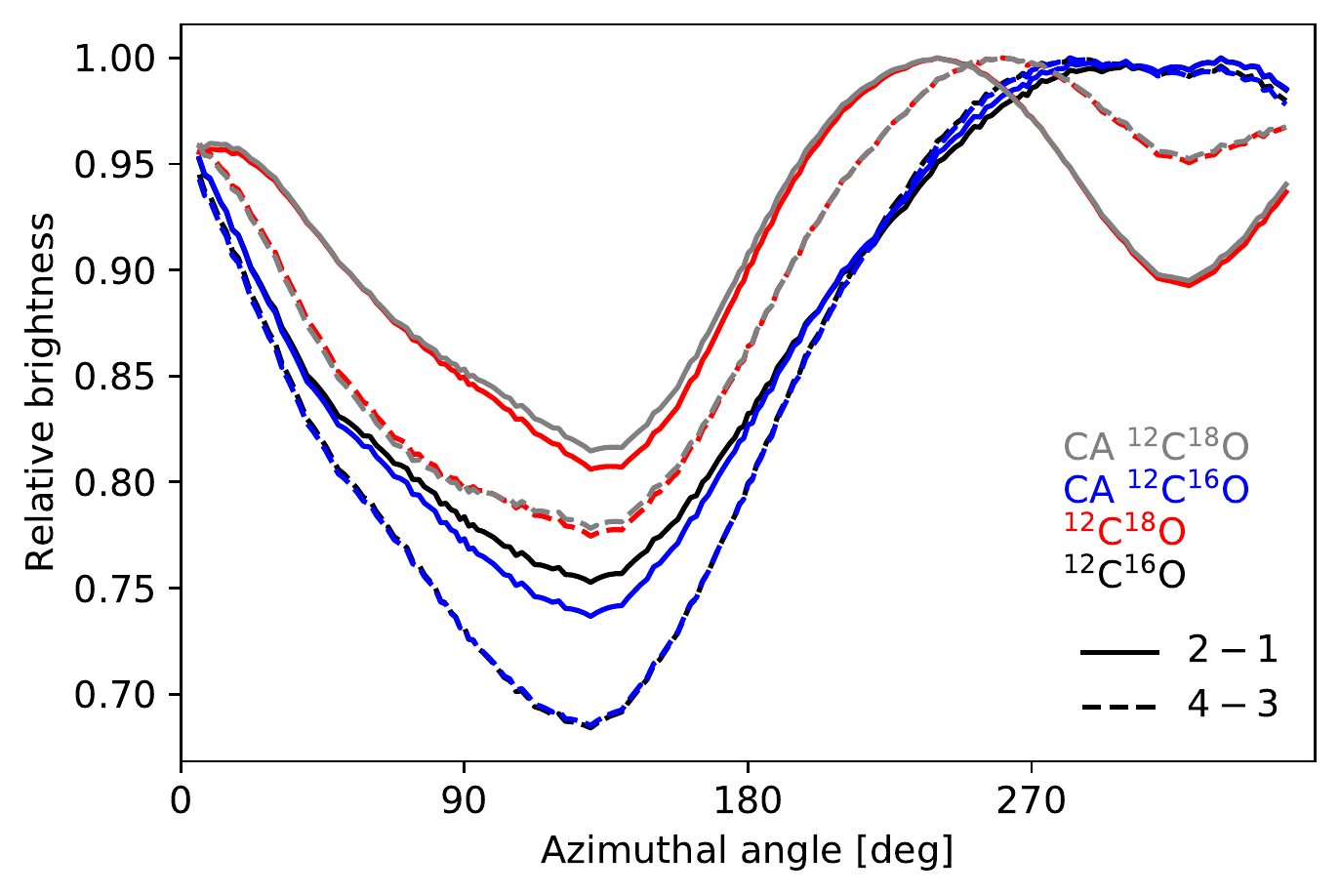}% made with azbrightness.ipynb
\label{fig:azc}
}
\qquad
\subfloat[d][]{
\includegraphics[width=0.43\textwidth]{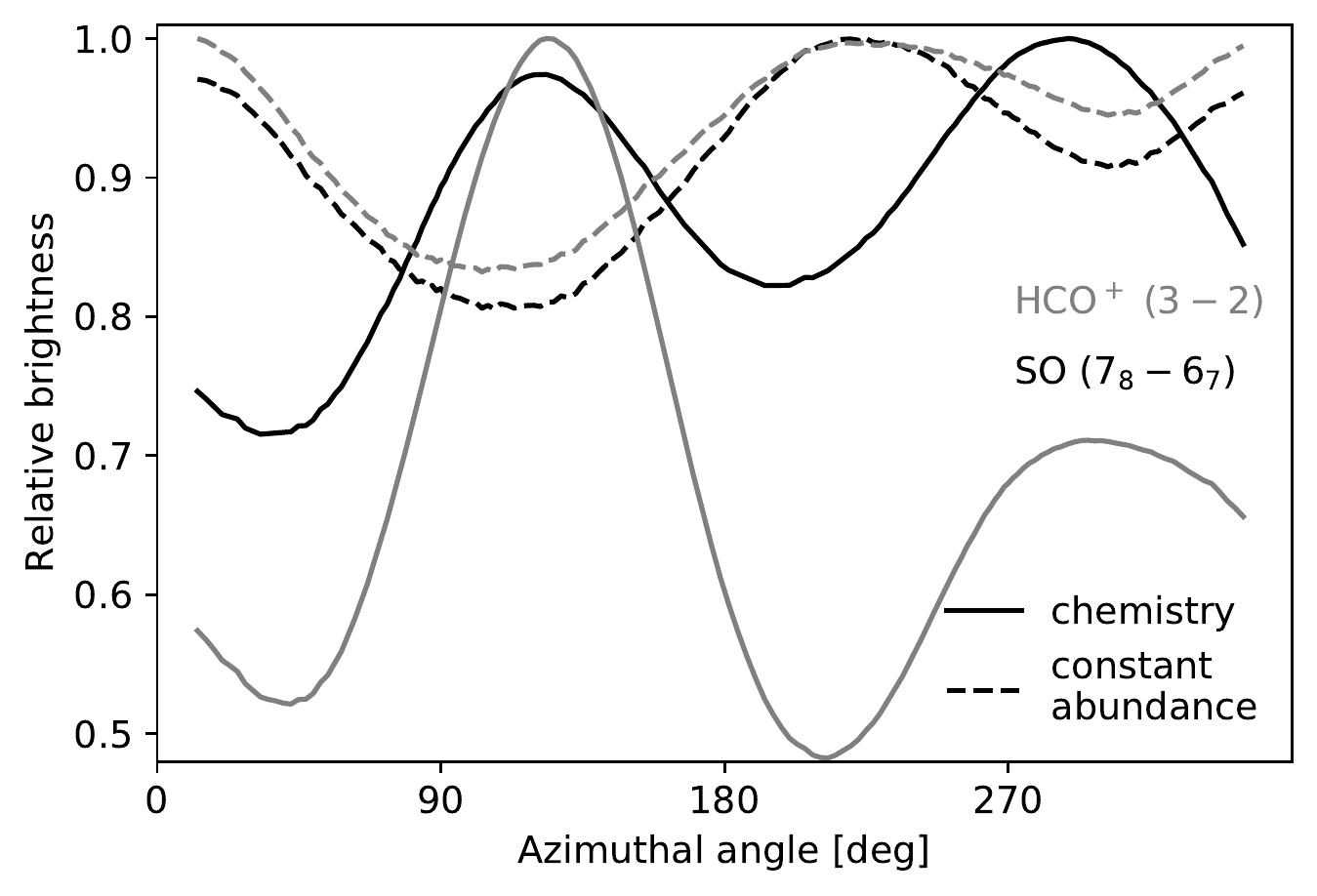}
\label{fig:azd}
}

\caption{\referee Comparison of the azimuthal brightness profiles of integrated intensity for annuli of $28<r<32$~au  ($r\sim0.5$~arcsec). (a) $^{13}$CO~$(2-1)$ moment 0 maps of the misaligned and aligned discs at a 30$^{\circ}$ inclination with the disc in the original position ($\phi=0$) and rotated before inclining ($\phi=90$). The image is deprojected to extract the annulus. (b) Three CO isotopologues for the misaligned disc at $i=0^{\circ}$. (c) Radiative transfer models using a constant CO abundance and abundances obtained from chemical modelling with $i=0^{\circ}$. The azimuthal variation traces the temperature variation rather than variation in abundance. The higher transitions are most sensitive to temperature and so the profiles are most similar when a constant abundance is used. (d) Comparison of azimuthal profiles for HCO$^+$ and SO at $i=0^{\circ}$ with calculated abundances and fixed abundances.}
\label{fig:azbright}

\end{figure*}

}

Higher molecular energy levels are excited at higher temperatures and densities so we might expect different transitions to trace different features.  Fig.~\ref{fig:azb} shows the azimuthal brightness profile of the (face-on) moment 0 maps for three CO isotopologues and two different transitions, normalised by the maximum value of each profile. We see the azimuthal variation increasing for the higher transitions due to their increased sensitivity to the temperature variation. In all three transitions there is clearly a `hot' side and a `cold' side to the disc. The more abundant isotopologues are also more sensitive to the temperature variation because they are more optically thick and trace gas higher in the disc where the temperature variation is greater.

{\referee
There are two effects here that contribute to the azimuthal variation in line emission. To distinguish between the contributions of the variation in excitation level due to the temperature asymmetry and the molecular column density we compare azimuthal profiles of line emission with the calculated abundances to those modelled with constant molecular abundances in Figs.~\ref{fig:azc} and \ref{fig:azd}. We find only a small azimuthal variation in CO column density (Figs.~\ref{fig:coldensslices} and \ref{fig:coldensspatial}) and we obtain a very similar azimuthal variation in integrated intensity with a constant CO abundance as with calculated abundances (Fig.~\ref{fig:azc}), which indicates that the dominant effect in this case is due to the gas temperature rather than molecular column density. For HCO$^+$ however, the azimuthal variation in integrated intensity significantly increased by the variation in column density (Fig.~\ref{fig:azd}). The HCO$^+$ emission varies azimuthally by $\sim 15$ per cent due to the temperature variations alone. This increases to $\sim 50$ per cent when accounting for chemical changes in response to the physical conditions. The variation in SO emission is greater when the abundances are taken from the chemical calculation but to a lesser extent than for HCO$^+$. There is also an azimuthal offset in the peak of the SO emission between the chemical model and the constant abundance model. 
}

{\referee
\section{Validity of the simplified static model}
\label{sec:timescales}
We have assumed a simple model in which temperatures and chemical abundances are calculated from a static snapshot as post--processes from a purely hydrodynamical model. In reality, the chemistry is coupled to the dynamical and thermal evolution of the disc. We now examine the validity and limitations of approximating the system in this way. By assuming radiative equilibrium, any temperature lag due to finite cooling times is not taken into account. First, we consider the thermal timescales in the disc and the validity of assuming radiative equilibrium. Similarly, we assume that the chemical abundances have reached a steady state. However, the abundances of various molecules evolve on different timescales. We therefore examine the chemical timescales and consider how this might affect the results.

\subsection{Thermal timescales}
For the assumption of radiative equilibrium to be valid the cooling timescale need to be shorter than the dynamical timescale, which in this case the orbital period of the disc at a given radius. We consider two key factors affecting the cooling timescale, namely the dust cooling timescale and the thermal equilibrium timescale for gas and dust.

\subsubsection{Dust cooling timescale}

We estimate cooling timescales assuming that the most efficient cooling mechanism in the region of the protoplanetary discs studied here is via dust emission and following \citet{chiang1997}. 

The internal energy per unit area is
\begin{equation}
U=\frac{1}{\gamma-1} k_{\mathrm B} \frac{\Sigma}{\mu m_{\mathrm p}} T_{\mathrm c},
\label{eq:intenergy}
\end{equation}

where $\gamma = 1.4$ is the adiabatic index, $k_{\mathrm B}$ is the Boltzmann constant, $\Sigma$ is the mass surface density, $\mu m_{\mathrm p} = 2.3 m_{\mathrm p}$ is the mean molecular mass and $T_{\mathrm c}$ is the temperature in the bulk of the disc, for which we take the temperature at the midplane. The cooling timescale is then

\begin{equation}
t_{\mathrm {cool}} \approx \frac{U}{2\sigma T_{\mathrm {surf}}^4},
\label{eq:tcool}
\end{equation}

where $\sigma$ is the Stefan-Boltzmann constant and $T_{\mathrm {surf}}$ is the dust temperature at the surface of the disc. We assume that the dust and gas temperatures are well--coupled (which will be discussed below). Taking values from the warped disc model, we find that the cooling timescale is far shorter than the orbital period, $P$. At $r=10$~au, $t_{\mathrm {cool}}\sim 23$~days and $P\sim32$~years and at $r=30$~au, $t_{\mathrm{ cool}}\sim 240$~days and $P\sim165$~years. 

The temperature of material moving in or out of a shadow will therefore quickly adjust within a small fraction of its orbital period and we would expect to see temperature differences between shadowed and illuminated parts of a disc. The inner disc precesses at the beginning of the simulation but aligns before completing a full precession period, within a few thousand years. We can therefore consider the position of the inner disc as effectively fixed and the shadows that it casts to be stationary.

%%% Thermal coupling of gas and dust.
\subsubsection{Thermal equilibrium time-scales for gas and dust}
We assume that the dust and gas are in thermal equilibrium and that the gas is cooled by its thermal coupling to the dust. For this assumption to be valid, the timescale for the gas and dust to become thermalised must be shorter than the dynamical timescale. We now examine the timescale for the gas to thermalise with the dust.

The energy transfer rate between the dust and gas found by \citet{youngk2004} is
\begin{equation}
\begin{aligned}
\Lambda_{\mathrm{gd}} = {} & 9.0\times10^{-34} n(\mathrm H)^2 T_K^{0.5}\left[1-0.8\exp\left(\frac{-75}{T_K}\right)\right] (T_K-T_d) \\
 & \left(\frac{\Sigma_d}{6.09\times10^{-22}}\right) \mathrm{ergs~cm}^{-3}\mathrm{s}^{-1}.
\end{aligned}
\end{equation}
$T_K$ and $T_d$ are the gas and dust temperatures respectively and $\Sigma_d$ is the average dust cross--section per baryon. With the dust parameters of our model we calculate $\Sigma_d = 1.26\times10^{-21}$~cm$^{-2}$. Assuming the dust temperature responds instantly to the change in irradiation as it orbits in the disc and that the gas temperature is slower to respond and is still the same as the temperature of the dust on the opposite site of the disc, we set $T_d = T_d(r,z)$ and $T_K = T_d(r,-z)$. 

\begin{figure}
\centering
\includegraphics[width=6cm]{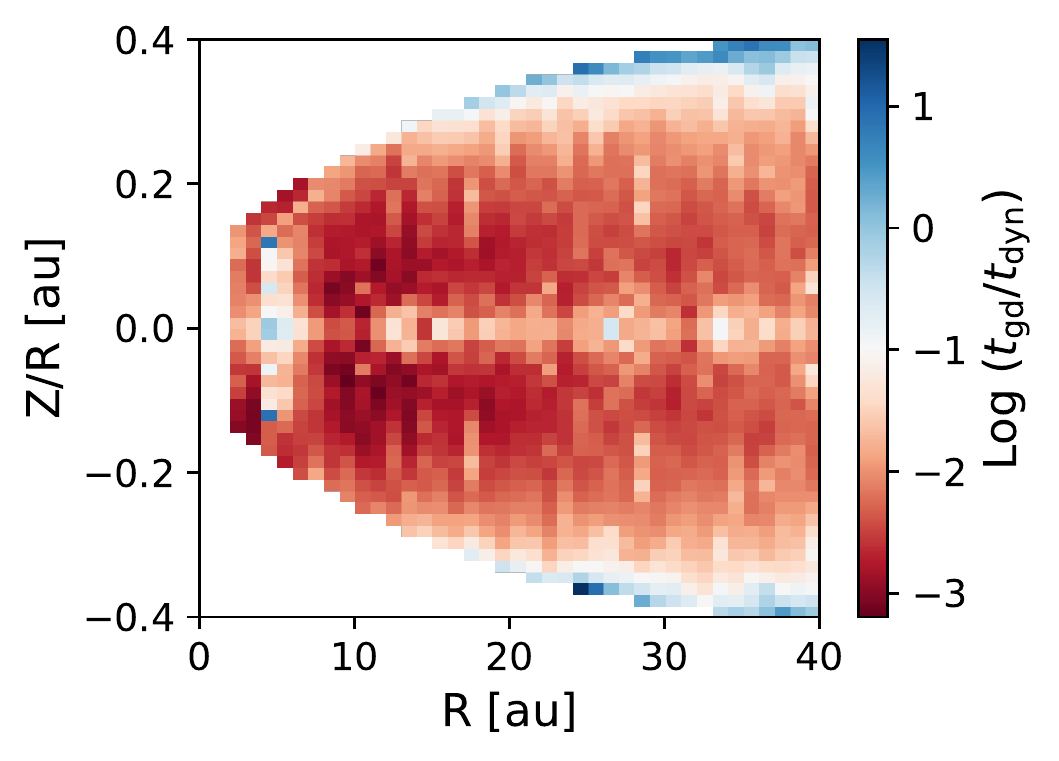}% made with slice_timescales.py
\caption{Estimated ratio of the timescale for the gas and dust to be thermalised ($t_{\mathrm{gd}}$) to the orbital period ($t_{\mathrm{dyn}}$). If $t_{\mathrm{gd}}$ is a small fraction of the orbital period the assumption of local thermodynamic equilibrium is valid. Regions with $t_{\mathrm{gd}}/t_{\mathrm{dyn}} > 0.1$ are coloured blue and this assumption does not hold. }
\label{fig:thermkep}
\end{figure}

We estimate the timescales for the gas and dust to become thermalised for a slice of the disc taken through the warp. Fig.~\ref{fig:thermkep} shows the resulting ratio of the timescale for the gas and dust to become thermalised to the Keplerian orbital period. In regions coloured red $t_{\mathrm{gd}}/t_{\mathrm{dyn}} < 0.1 $ and we see that the assumption of local thermodynamic equilibrium holds in nearly all of the modelled region. Note that the gap is located at $\sim$5~au and the density there is very low. The gas and dust are poorly coupled only in the very top model cells and the dominant cooling mechanism is likely to be molecular line emission in those regions.  
}

{\referee
\subsection{Chemical timescales}
\label{sec:mainchemtimes}

Here we consider the relevant timescales for the formation and destruction of key molecular species. The chemical model is evolved from ISM abundances for 1~Myr for fixed physical conditions and we assume that steady state has been reached. However, we expect the abundances to reach steady state more quickly starting from abundances calculated from the physical conditions of the disc than from ISM abundances. If the reaction timescales governing a molecule are shorter than the dynamical timescale (i.e. less than the orbital period at a given radius) then we expect its abundance to change due to azimuthally varying physical conditions. We now examine the time evolution of the chemical abundances of selected cells to assess the suitability of this assumption.

We compare the evolution for selected cells on opposite sides of the disc, on the `bright' and `dark' sides. To do this, slices are taken along the $y$-axis and the physical values of the SPH particles (centres of the Voronoi cells) are interpolated onto a regular grid. This way, it is possible to extract the abundances at equal radii and heights on opposite sides of the disc. In the outer disc, $z\prime \approx z$ because the disc is only slightly misaligned with respect to the X-Y plane outside of the planet's orbit. The cells we select are in regions where we find the abundances to differ most on opposite sides of the disc.

First we determine whether chemical abundances have reached a steady state when they are extracted at 1~Myr. In Fig.~\ref{fig:ISMchemevo} we plot the chemical evolution from pre--calculated ISM abundances. After 1~Myr, the key species are close to steady state. An exception is CO on the shaded side of the disc, for which freeze out accelerates after $10^4$ years. Gas phase SO is also substantially depleted on the shaded side, with negligible abundance in much of the disc. More details on the chemical timescales and the steady state assumption can be found in Appendix~\ref{sec:chemtimes}.

\begin{figure}
\centering
\includegraphics[width=0.9\columnwidth]{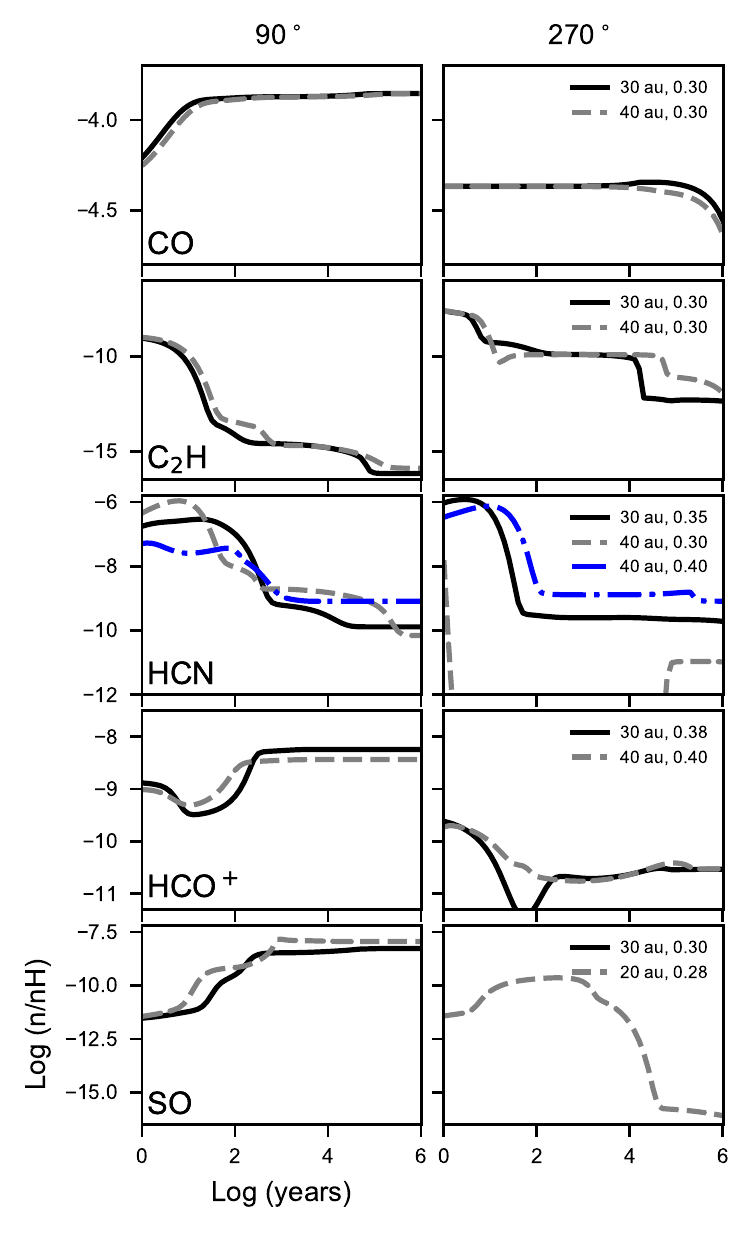}
\caption{Evolution of the chemical abundances from ISM abundances of selected cells on the positive $y$-axis (90$^{\circ}$, {\refereeB the bright side}) and negative $y$-axis (270$^{\circ}${\refereeB, the shaded side}) as defined in Fig.~\ref{fig:coldensXY}. Radius and $z/r$ coordinates of the cells are given in the legends.}
\label{fig:ISMchemevo}
\end{figure}

\begin{figure}
\centering
\includegraphics[width=0.9\columnwidth]{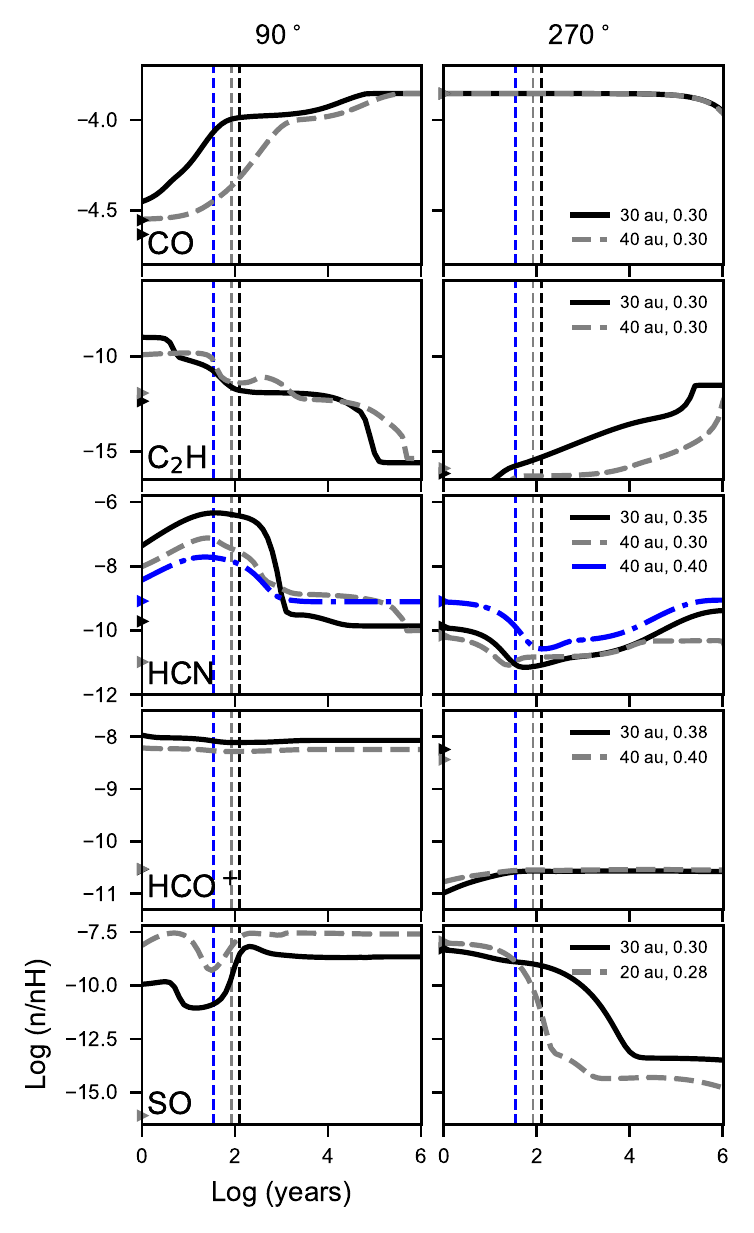}
\caption{Evolution of the chemical abundances of selected cells as for Fig.~\ref{fig:ISMchemevo}. Blue, grey and black dashed lines indicate the dynamical time $t_{\mathrm{dyn}} = t_{\mathrm{orbit}}/2$ at 20, 30 and 40~au. Triangles indicate the initial abundances for cells with the colour scheme shown in the legends. The initial abundance of SO at $r=$20~au, $z/r=0.28$ is negligible and not shown.}
\label{fig:discchemevo}
\end{figure}

Next, we examine whether the chemical abundances will evolve over the course of an orbit like predicted by our simple static model. Chemical evolution depends on the initial abundances. In our model, we neglect the disc formation phase and abundances in the disc are calculated starting from ISM abundances. However for this test, we initialise the chemical model with abundances that were calculated for the disc to obtain a more accurate approximation of the chemical evolution over an orbit. 
For the same cells as above, we initialise the chemical model for each cell with abundances from the equivalent cell on the opposite side of the disc, which were extracted from the former model after 1~Myr. 
%These abundances are used as input for the cells at the same $z/r$ on the opposite side of the disc.
The chemistry is then evolved under the physical conditions of the cell. In Fig.~\ref{fig:discchemevo}, the "$90^{\circ}$" ("$270^{\circ}$") panels show the chemical evolution on the brighter (shaded) side of the disc with initial abundances taken from the equivalent cell at $270^{\circ}$ ($90^{\circ}$) after 1~Myr which are shown in Fig~\ref{fig:ISMchemevo}.

HCO$^+$ rapidly reaches steady state on both sides of the disc, indicating that the abundances respond very quickly to changing physical conditions and are accurately described by the steady state assumption. The other key species do not always reach steady state within the dynamical timescale which means that the exact nature of the azimuthal variation is likely to be more complex than that found in static, steady state calculations.

The CO evolution is highly dependent on the initial abundance. Freeze-out only begins to deplete gas--phase CO after around $10^4$ years in this model so the gas--phase CO abundance would not show azimuthal variation. However, this does not affect our conclusions regarding the asymmetry in CO emission because the temperature difference rather than abundance is the dominant factor. A larger warp with a greater azimuthal temperature contrast may impact the location of the CO snowline.

HCN appears to have reached steady state by 1~Myr. However, HCN abundances peak on the bright side of the disc on a dynamical timescale. The steady state abundances are between a factor of 10-1000 lower than the abundances after a dynamical timescale on the bright side and a factor of 10-100 higher on the shaded side (see Fig.~\ref{fig:discchemevo}). We expect the asymmetry in HCN abundance to be much greater in a fully coupled chemical--dynamical model.

Since SO has not reached a steady state on the cooler side of the disc the depletion in that region is overestimated. Additionally, in Fig.~\ref{fig:discchemevo}, we see the SO abundance decrease where the HCN peaks. This suggests that the azimuthal contrast would be less pronounced than predicted in the simplified model.

C$_2$H reaches steady state by 1~Myr and similar abundances are reached on this timescale from ISM and disc initial abundances. The steady state abundances are higher on the cooler side of the disc however the opposite is true on dynamical timescales due to the differing rates on opposite sides of the disc and the sensitivity to the initial abundances. C$_2$H abundances are likely to be around three orders of magnitude higher in a dynamical chemical model since the steady state timescale is much longer than a dynamical timescale.

%%%%%

In summary, the key species reach steady state by the time abundances are extracted on the warmer side of the disc at 1~Myr. On the cooler side, CO and SO abundances are still decreasing at this time. Next, we compared the chemical and dynamical timescales. Some species such as HCO$^+$ respond quickly to the change in physical conditions. Other species do not reach a steady state within the dynamical timescale when evolving between the warmer and cooler sides of the disc. Consequently, for those species the asymmetry in abundance may be greater in a coupled chemical--hydrodynamical model, such as for HCN, or the abundance may be more symmetrical, as we predict for SO. Alternatively, the asymmetry in the abundance may be offset from the asymmetry in physical conditions due to a time lag.
}

\section{Discussion}
Just as scattered light morphology is affected by azimuthally asymmetric illumination, so are the chemical abundances and line emission. Features observed in line emission can be related to the warped structure of a disc and the presence of a misaligned inner disc, which provides another probe of disc structure. In this section we examine how the effects of a warped protoplanetary disc structure on the chemistry and line emission may be useful for identifying such discs.

{ \referee 
\subsection{Best tracers of warp structure}

The easiest molecule to detect in protoplanetary discs is CO and, since the moment 0 maps for $^{13}$CO transitions show a 20-30\% azimuthal variation in brightness (see Fig.~\ref{fig:azbright}), this promises to be a useful tracer of warps. This variation in intensity occurs at radii $>30$ au, corresponding to a separation of $>0.2$~arcsec at a nominal distance to protoplanetary discs of 140~pc. ALMA observations are therefore capable of detecting this structure in the majority of discs. HCO$^+$ looks to be a good indicator of warped structure due to the strong asymmetry present even at a moderate inclination.

SO lines are fainter and the morphology is likely to be more difficult to interpret. The variation in appearance with azimuthal position of the disc structures may open up possibilities for placing tighter constraints on the orientation and structure of the disc, particulary when analysed alongside scattered light images.

Given the difficulty of detecting many molecular species in protoplanetary discs, the primary concern when selecting the transition to observe is its brightness. Although transitions that are more sensitive to temperature will show more azimuthal variation, the variation is also very likely to be detectable in lower transitions so it is probably more beneficial to observe the brightest line for a given species.

The best tracer of this warped disc structure also depends upon the resolution and scale of the observations. We saw in Fig.~\ref{fig:mom0maps} that the asymmetry in $^{13}$CO~$(4-3)$ emission occurs at $r>30$~au. This means that the structure is potentially detectable in more distant discs or with lower resolution observations. Since SO is faint, the brighter clumps alone are more likely to be detected than the more extended `bow-tie' structure. The SO clumps would require high angular resolution and a high sensitivity to detect.

This study considers a small warp due to a planet at 5~au and we expected the strongest effects to be present in line emission on smaller scales $<50$~au. It appears that a planet at 5~au can be responsible for asymmetries in the column density of molecules such as CN and C$_2$H at much larger radii, beyond the region of the disc we have modelled. We propose that asymmetry in these species may be a useful signpost of a warped disc but modelling of a larger disc is required to verify this. In addition, the static chemical model implemented here is likely to have underestimated the abundances of HCN, CN and C$_2$H. We expect these molecules to show some azimuthal variation but a dynamical chemical model would be necessary to explore their behaviour accurately.
}

\subsection{Distinguishing the effects due to a warp from other scenarios}

There are other structures and mechanisms that can result in asymmetry in line observations but it should be possible to distinguish a warped structure from these scenarios.  We now consider a few possible scenarios that could cause asymmetrical line emission.

 An unresolved outflow or disc wind would have a very different velocity signature that would be inconsistent with Keplerian or near--Keplerian rotation. A spiral structure in the disc would have a distinct morphology and is likely to be visible in continuum emission {\referee where it is coincident with the dust disc. The clumps of emission seen in SO and HCO$^+$ could be explained by physical clumps of material within the disc.} Again, these would have clear counterparts in the submillimetre continuum if they were due to physical density structures.

Another possible source of asymmetry is asymmetric external irradiation by a nearby star. This too can be verified since the properties and positions of nearby stars are known. The viewing inclination may also be responsible for asymmetry in moment 0 maps. 
%We see this, for example, in the $^{13}$CO~$(4-3)$ maps in Fig.~\ref{fig:alignmom0}. 
{\referee The effects of inclination can be understood through modelling, as we have explored. We have shown that there are differences in the azimuthal profile of CO emission between a warped disc and unperturbed disc at an inclination of $i=30^\circ$.}

As discussed earlier, the abundance of HCO$^+$ is sensitive to the X-ray irradiation. Variations in the height of the inner disc due to a planet--induced spiral, for example, may cause shadowing that could result in asymmetric emission similar to the morphology we expect due to the warp and misaligned inner disc. For this reason, temperature--dependent lines could provide more robust tracers of warps than those very sensitive to ionization because they are affected less by an uneven disc surface.

Lastly, from an observational perspective, asymmetries can arise from foreground contamination. A clear example of this is the AS~209 disc. In the $^{12}$CO~$(2-1)$ moment 0 map, one side of the disc is clearly brighter than the other but this is due to cloud contamination \citep{guzman2018}. In addition, asymmetries can be introduced into line observations when continuum subtraction is performed at a location in the disc where there is a bright dust arc in the continuum \citep{vanderplas2014,boehler2017}. 

\subsection{Considerations for real systems}
{\referee
In this paper we study one possible disc configuration and a single set of stellar properties. The morphology of the line emission will vary for different systems. We have also made a number of assumptions to simplify the model. We discuss additional parameters that will likely impact on how the chemistry and molecular line emission of the disc is affected by a warp below.}

\subsubsection{Star and disc properties}
We chose to model a system in which the orbital plane of the planet are the disc are only slightly misaligned. This has shown that, even for small misalignments, the chemical abundances in the disc are affected. A greater misalignment gives rise to narrow shadows cast by the inner disc on the outer disc instead of a single broad shadow {\referee (e.g. \citealt{min2017,facchini2018,nealon2020b})}. The accompanying warp will be more pronounced, with a greater azimuthal variation in temperature. The mass of the planet also affects the amplitude of the warp and therefore the magnitude of the azimuthal temperature variation.

The inner disc and warped outer disc have different precession times, therefore the relative positions of the warp and inner disc are variable. In this case though, the inner disc has aligned with the orbit of the planet. Both the misaligned inner disc and warp cast shadows and contribute to temperature variations in the outer disc and the extent of the disc that is shadowed will be greater if the inner disc tilt is not aligned with the warp. The implication is that if the inner disc and warp are in different relative positions to that studied here a greater section of the disc will be subject to reduced irradiation, which will affect the chemistry. The CO distribution, for example, is affected mainly by temperature variations, whereas lines sensitive to ionization are more sensitive to the orientation of the inner disc since this has the dominant effect on X-ray irradiation and shadowing.

The total luminosity and X-ray luminosity will also affect the morphology of the line emission and the brightness of molecular lines. A hotter central star would result in a warmer disc with more distant snowlines. The other factor to consider here is the distance of the planet from the star. A warp further out in the disc would intersect with different snowlines. The 10-20~K variation in temperature at the location of the warp may cross the freezeout temperature for other species leading to alternative tracers of the warp structure.

We have briefly examined the effect of the inclination of the system relative to the observer and shown that this affects the morphology of the emission for many lines. It is complicated further by the nonaxisymmetry of the system, meaning that the azimuthal position of the structures in the disc also affects the observed emission. At higher inclinations dust shadowing should also be considered.

We have not considered the dust distribution in this model but assumed a constant gas-to-dust ratio throughout the disc and a constant power law size distribution. An observational effect that has been encountered is that the continuum subtraction may falsely create asymmetry in the line maps where there is an enhancement such as an arc in the continuum due to a dust trap. Additionally, the grain properties affect the chemistry. For example, the dust size distribution affects the rate of freezeout \citep{cleeves2016} and hence the gas phase CO column density. The deposition of volatiles is quicker onto ice than onto bare grains so any molecular depletion due to freezeout could be more dramatic further out in the disc or, for species with higher freezeout temperatures, where there is already an ice layer on the dust.

We modelled a compact (50 au) disc for computational reasons but CN, for example, increases in relative abundance at larger radii. Such species could therefore trace asymmetries on larger scales that are a direct result of the structure in the central $r<10$~au but we are currently unable to verify this.

{\referee Another factor to consider is that there will be some vertical mixing of gases. The mixing timescale will depend upon the turbulence in the disc, which is known to vary between discs. This process may act to increase any azimuthal asymmetry in abundances if gas is advected vertically from regions where the chemical timescales are shorter. The result would be increased variation in abundances in disc layers where the reaction rates are much slower.}

We have studied one particular scenario and, as discussed above, the distribution of molecular abundances and the spatial variation in line emission will be affected by many factors. The exact effects of the warp and shadowing on molecular line emission will vary for different systems, therefore a model will have to be constructed for the properties of the individual observed system to explain the observations. 

\subsubsection{Equilibrium assumptions}
{ \referee
% casassus cooling lag
The radiative equilibrium calculations implicitly assume that the gas and dust are perfectly coupled and that the dust cools instantly. This provides a useful starting point for studying possible effects of a warped disc structure but the reality is, as always, more nuanced. \citet{casassus2019} presented a study of temperature changes of disc material as it passes into shadow. They confirm that there is a thermal lag and that the temperature decrement in a shadowed region depends on the surface density and other properties of the disc. The radiative equilibrium assumption may be considered as a limiting case and the differences in molecular column density as a maximum limit. Furthermore, not all shadowed discs show temperature drops \citep{casassus2019}. If temperature drops depend on the disc properties, then so will any variations in molecular column density due to shadows. However, it is possible that molecular species produced by FUV or X-ray driven reactions may still show an azimuthal variation that is independent of the temperature decrement.

% HCN and SO near critical density
We assume LTE for the molecular line transfer calculations and this is valid for the inner disc region studied here. The only exception to this is the narrow low density region above $z/r = 0.3 $ for the HCN $(3-2)$ transition for which the critical density ($n_c \sim 6.5\times10^7$~cm$^{-3}$) is not exceeded. Non-LTE treatment would be required to consider radii $>40$~au since the density of the molecular layer would fall below the critical density of many molecular transitions of interest.

\subsubsection{Chemical modelling}
The method of performing the chemical calculation on a single snapshot from the hydrodynamical limits the accuracy of the calculated abundances for species driven by reactions with timescales longer than a dynamical timescale. We have highlighted which species are affected and discussed how this affects the results. A full dynamical chemistry calculation for this disc model would be prohibitively computationally expensive (adding of order $\sim 10^4$ cpu hours for each hydrodynamical time step) and this would still not produce generic results due to the sensitivity of the results to the exact disc morphology, disc size, radiation field and assumed initial abundances. The model could be simplified, for example, if certain species of interest are identified for a system. A reduced chemical network that is known to reproduce the evolution of those species adequately could be used. Alternatively, the chemical calculation can be restricted to an annulus where an asymmetry in physical conditions has been identified or asymmetric molecular emission has been detected.
}

%X-ray effects \& \cite{arulanantham2018} FUV effect on CN and HCN
\subsection{Comparisons to observed discs}
In this section, we look for signs of asymmetry in published molecular line observations. We do not expect to see identical signatures to our model line maps because the host stars have different luminosities and disc sizes and the structures are different.

{\bf HD142527:} 
%. Analysis of the kinematics indicates that the disc is likely to be warped \citep{rosenfeld2014}. 
This is a circumbinary disc with a large ($r=100$~au) cavity. A low mass companion was discovered at  $r\sim13$~au \citep{biller2012,close2014}. The HCN $(4-3)$ and CS $(7-6)$ emission forms a ring at $r\sim 220$~au that is azimuthally asymmetric \citep{vanderplas2014}. This is attributed to the arc seen in the dust continuum emission either through lower temperatures caused by the enhanced cooling due to the flatter dust distribution or because the optical depth is higher in the dusty arc. A third possibility is that there is asymmetric illumination due to inner disc shadowing which drives a difference in the chemistry. The HCO$^+$ $(4-3)$ emission forms a ring with a brighter arc on one side. {\referee \citet{casassus2013} argue that variations in the stellar heating can be ruled out as the cause in this case because this emission forms a ring rather than a broken arc. This indicates that the disc has a direct line off sight to the star.}

Hydrodynamical models matching the parameters of the system demonstrate that interaction between the companion and the disc gives rise to all of the observed structures \citep{price2018ab}. A transient misaligned circumprimary disc casts shadows onto the outer disc. The shadows are narrow due to the large misalignment and so the effect is not directly comparable to the models presented here of a small misalignment with a broad shadow. Nevertheless, $^{13}$CO $(2-1)$ and C$^{18}$O $(2-1)$ emission maps show a nonaxisymmetric ring \citep{boehler2017,price2018ab} with breaks at the approximate locations of the narrow shadows. This is a feature we would expect to see as a result of the lower temperatures in the shadows increasing the freezeout rate of CO as well as reducing the emission. {\referee \citet{yen2020} show there is a pressure bump acting as a dust trap on the northern side of the disc and this could be the cause of the arcs seen in CS, HCN and HCO$^+$ but the cause of the asymmetric CO emission is still unclear.}

{\bf2MASS J16042165-2130284: } 
This system is believed to have a strongly misaligned inner disc which casts narrow shadows on the outer disc \citep{mayama2012}. These shadows moved between observations three years apart \citep{pinilla2018} and may appear to `rock' back and forth across the same sides of the disc \citep{nealon2020b}. Again, the misalignment is considerably greater than the scenario modelled in this paper and we would not expect to see reduced brightness or shadows across large sections of the disc. {\referee The HCO$^+$~$(4-3)$ \citep{mayama2018} and CO~$(3-2)$ \citep{mathews2012} integrated intensity maps reveal dips on opposite sides of the disc which coincide with those seen in the near-IR image. We find that HCO$^+$ abundances respond rapidly to shadowing, therefore it is plausible that the dips in HCO$^+$ emission are due to reduced abundance caused by the shadowing of stellar ionizing radiation. Variation in HCO$^+$ emission has been observed on the timescale of days \citep{cleeves2017} which supports the idea that the reaction rates are fast enough to result in reduced HCO$^+$ abundance in the shadows. For CO however, we find that asymmetry is mostly due to the underlying temperature structure rather than a change in abundance. This suggests that the narrow dips are indicative of a temperature drop but detailed analysis of the expected cooling time in this disc would be necessary to determine whether this is the case.}

{\bf HD143006:} %Giulia's paper narrowlane  shadows %CO is depleted in the inner disc, 
The scattered light image of this system shows a striking brightness asymmetry between the East and West sides of the disc \citep{benisty2018}. This is thought to be the result of shadowing by a moderately inclined inner disc with misalignment of either $8^{\circ}$ or $24^{\circ}$ \citep{perez2018}. The CO emission has the same East-West asymmetry as the scattered light. The optically thick CO~$(2-1)$ traces temperature structures so this is what we would expect to see for shadowed region. {\referee \citet{perez2018} and }\citet{pinte2020} report a `velocity kink' in the CO channel map that could indicate the presence of a planet in the gap in the dust rings between the inner and outer discs. Observations of a more optically thin tracer like C$^{18}$O would be informative to help determine whether the lower temperature results in CO depletion as we predict. 

{\bf TW Hydrae:} 
This system is the most similar to our model since there is a broad shadow across the disc causing the disc to be fainter to the North-West in scattered light \citep{vanboekel2017}.There may be hints of asymmetry in $^{12}$CO in the inner 1 arcsec \citep{teague2019}. No significant azimuthal variation was found in N$_2$H$^+$, C$_2$H or CS emission  \citep{qi2013,kastner2015,teague2017} but these observations cover larger scales than we have modelled. High angular resolution molecular line observations are required to determine whether there is any azimuthal variation on subarcsecond scales as we would expect if there is an inner disc with a small misalignment.

{\bf HD~100546:}
This Herbig Ae star hosts a disc with a $r=14$~au cavity. The system is therefore not directly comparable to the model presented here but we would still expect asymmetries in line emission if there is a warp and/or misaligned inner disc as suggested by \citet{walsh2017}. There are signs of structure in CO $(3-2)$ emission which may trace a two--armed spiral \citep{pineda2019} and SO emission has been detected on one side of the disc, coincident with a hot spot in CO \citep{booth2018}. The SO emission was proposed to trace a disc wind, accretion shock onto a protoplanet or a disc warp. The elongated morphology seen in the SO integrated intensity is similar to the structure we see in the inclined model emission maps and so it is possible that the SO traces a warped structure which occurs on a larger scale in HD~100546 than in the model presented here. {\referee In high resolution maps the western side of the disc is brighter in CO \citep{perez2020} which could be due to asymmetric illumination due to a misaligned inner disc or warp. Indeed the moment 1 maps show a strong deviation from Keplerian rotation towards the location of the inner disc which indicates misalignment. }

\section{Conclusion}

We have shown that the azimuthal temperature variation in a warped disc with additional shadowing from a misaligned inner disc leads to nonaxisymmetric column densities of many molecular species. This is true for the case studied here with only a small (12$^{\circ}$) misalignment between the planet's orbit and the outer disc. Therefore, azimuthal chemical effects should be considered when interpreting line maps of systems that may be warped and/or shadowed.

The azimuthal variation in molecular column density and temperature leads to measurable asymmetries in integrated intensity maps of the molecular emission and we show that this is true even for a disc observed at a moderate inclination. {\referee The best tracers were found to be CO, HCO$^+$ and SO on the scale of the disc considered here ($r<50$~au), with higher transitions being more sensitive to the azimuthal temperature variation. HCO$^+$ emission was the most strongly affected by the variation in abundance. A key result is that there is evidence of perturbation due to a planet on much larger scales than the planet's semimajor axis. The effect of the warp and misalignment on molecular line emission is a robust effect since we expect it to be seen in multiple tracers.}

The tracers we have identified are not necessarily universal because the chemical effects and line emission are not scale--free. For example, the relative positions of the warp and snow line affect observables. A warp at a different radius will intersect with the snow line of a different molecule, or none at all. In addition, the relative positions of warp and inner disc will change the azimuthal distributions of species because of the separate contributions to the temperature and irradiation variation due to shadowing by the inner disc and from the inherent shape of the warped outer disc.  {\referee The chemical timescale is similar to the dynamical timescale in parts of the disc for some molecules. Coupled chemical and hydrodynamical chemistry calculations are required to fully understand the behaviour of species such as HCN, C$_2$H and SO. For these reasons, a tailored chemo--dynamical model will be necessary for a given observed system to understand more fully the relationship between the disc structure and molecular line emission.}

%*Chemistry of planet formation*
There are hints of asymmetry in existing molecular line observations of protoplanetary discs. Recent  ALMA surveys have conducted molecular line observations of a large number of protoplanetary discs and most do not have counterpart scattered light observations that would easily reveal any shadows indicative of misaligned structures. Azimuthal variations in line emission may indicate the presence of a planet on misaligned orbit warping and breaking a protoplanetary disc. 
A broad sample of planet--hosting discs is required to test our understanding of planet formation models and planet--disc interactions. This provides a new probe of warped disc structures that will be useful for selecting disc targets that are likely to host a planet for detailed follow up observations. 

\section*{Acknowledgements}
The authors are grateful to the reviewer for their insightful and constructive feedback that helped improve the manuscript. This research used the ALICE High Performance Computing Facility at the University of Leicester and the DiRAC Data Intensive service at Leicester, operated by the University of Leicester IT Services, which forms part of the Science and Technology Facilities Council (STFC) DiRAC HPC Facility (www.dirac.ac.uk). The equipment was funded by BEIS capital funding via STFC capital grants ST/K000373/1 and ST/R002363/1 and STFC DiRAC Operations grant ST/R001014/1. DiRAC is part of the National e-Infrastructure. AKY, RA and RN gratefully acknowledge funding from the European Research Council (ERC) under the European Union's Horizon 2020 research and innovation programme (grant agreement No 681601). CW acknowledges financial support from the University of Leeds, the STFC and UK Research and Innovation(UKRI) (grant numbers ST/R000549/1, ST/T000287/1 and MR/T040726/1). RN acknowledges support from UKRI/EPSRC through a Stephen Hawking Fellowship (EP/T017287/1). Some figures in this paper were produced using SPLASH \citep{price2007}. We acknowledge the use of {\sc matplotlib} \citep{hunter2007}, {\sc numpy} \citep{harris2020} and {\sc astropy} \citep{astropy:2013,astropy:2018}.

\section*{Data availability statement}
Hydrodynamical simulations used the {\sc phantom} code which is  available  from \url{https://github.com/danieljprice/phantom}. The input files for generating the SPH simulations and radiative transfer models will be shared on reasonable request to the corresponding author. Radiative transfer calculations were performed using {\sc mcfost} which is available on a collaborative basis from CP. The chemical network and code are available on request to CW.



\bibliographystyle{mnras}
\bibliography{warpchembib} % if your bibtex file is called example.bib

%%%%%%%%%%%%%%%%%%%%%%%%%%%%%%%%%%%%%%%%%%%%%%%%%%

%%%%%%%%%%%%%%%%% APPENDICES %%%%%%%%%%%%%%%%%%%%%
\appendix
{\referee
\section{Chemical timescales and the assumption of steady state}
\label{sec:chemtimes}

The chemical calculations are evolved under the physical conditions of the disc for $10^6$ years, by which time we assume that a steady state has been reached. {\refereeB In other words, we assume that the chemical species of interest have had enough time to fully adjust to the physical conditions. A long chemical timescale indicates that a species is close to steady state. To verify this assumption, we examine the timescales required for the abundances to change further after the first $10^6$ years under the same constant physical conditions.} {\refereeB We note that this is different to estimating the timescales for the chemical abundances to adjust to a change in the physical environment. The chemical timescales are many orders of magnitude shorter earlier in the calculation, when the chemistry is far from steady state.} Fig.~\ref{fig:chemtimescales} shows the timescales for the formation/destruction of the key species after $10^6$ years for a sample of cells taken from a column through the disc at $r\approx 30$~au and $r\approx 40$~au. Considering a column in the disc gives an indication of how the timescales vary with height.

The timescales of most species are orders of magnitude longer than the dynamical timescale. This indicates that the abundances of those species have indeed stabilised. The abundance of CO, however, is changing on dynamical timescales in much of the lower (shaded) regions of the disc. In those regions, freezeout proceeds without reaching a steady state and the CO column density may be underestimated due to this excess freezeout. The molecular emission is affected less because $^{12}$CO and $^{13}$CO emission is optically thick and traces gas higher in the disc where the abundances are more stable.

\begin{figure}%plotted with plot_netrate_fullthickness_col.py
\centering
\includegraphics[width=0.9\columnwidth]{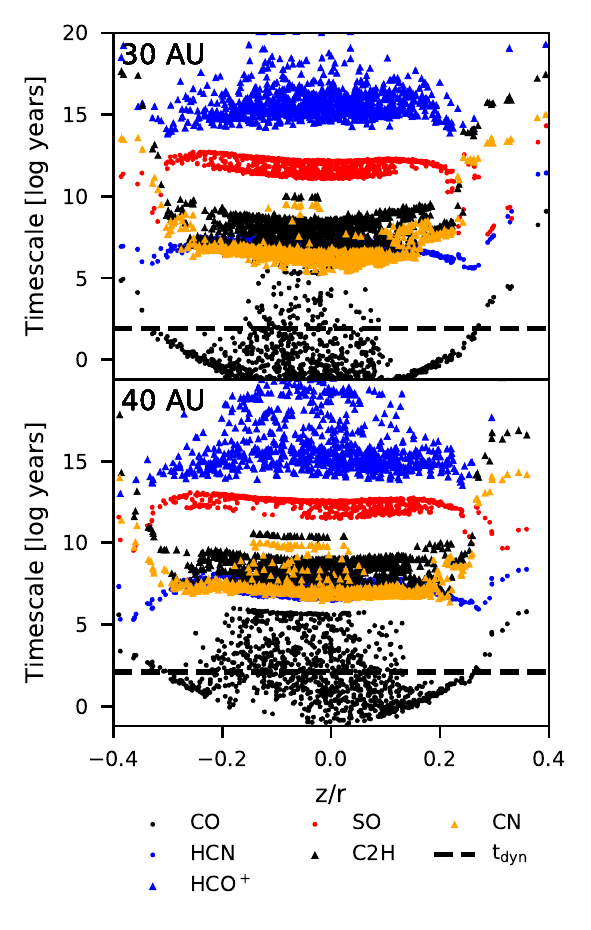}
\caption{Timescales for the formation/destruction of key species from a column through the disc at $x=0$, at $29.7 \mathrm{ au}<y<30.3$~au (top panel) and $39.6 \mathrm{ au}<y<40.4$~au (bottom panel). The dynamical time $t_{\mathrm{dyn}}=t_{\mathrm{orbit}}/2$ at $r=30$~au (top) and $r=30$~au (bottom) is indicated by the dashed lines.}
\label{fig:chemtimescales}
\end{figure}
}

% Don't change these lines
\bsp	% typesetting comment
\label{lastpage}
\end{document}